\documentclass[]{aastex63}
\usepackage{graphicx}
\usepackage{color}

\newcommand\blankline{\par\vskip 8pt\noindent}

\newcommand{\civres}{\ion{C}{4} $\lambda\lambda 1548,1552$}

\newcommand{\nvdoub}{\ion{N}{5} $\lambda\lambda 4604, 4620$}

\newcommand{\ciiinirtrip}{\ion{C}{3} $\lambda\lambda  9701 - 9719$}

\newcommand{\oiiising}{\ion{O}{3} $\lambda 5592$}
\newcommand{\oivdoub}{\ion{O}{4} $\lambda\lambda 3405 - 3414$}   %
\newcommand{\osixdoub}{\ion{O}{6} $\lambda\lambda 3811, 3834 $}   

\newcommand{\ovtrip}{\ion{O}{5} $\lambda\lambda 5571-5607 $}

\newcommand{\kms}{\hbox{km\,s$^{-1}$}}

\newcommand{\Msun}{\hbox{$M_\odot$}}
\newcommand{\Lsun}{\hbox{$L_\odot$}}
\newcommand{\Msunyr}{\hbox{$M_\odot \,\hbox{yr}^{-1}$}}
\newcommand{\Mdot}{\hbox{$\dot M$}}

\newcommand{\cmfgen}{\textsc{cmfgen}}

\shorttitle{Evolution of WC-Type Wolf-Rayet Stars in the LMC}
\shortauthors{Aadland et al.}
\graphicspath{{./}{figures/}}
\begin{document}

\title{The Physical Parameters of Four WC-type Wolf-Rayet Stars in the Large Magellanic Cloud: Evidence of Evolution\footnote{This paper includes data gathered with the 6.5 meter Magellan Telescopes located at Las Campanas Observatory, Chile.}}

\correspondingauthor{Erin Aadland}
\email{aadlander@lowell.edu}

%\author[0000-0002-0786-7307]{Greg J. Schwarz}
\author{Erin Aadland}
\affiliation{Department of Astronomy and Planetary Science, Northern Arizona University, Flagstaff, AZ, 86011-6010, USA}
\affiliation{Lowell Observatory, 1400 W Mars Hill Road, Flagstaff, AZ 86001, USA}
%\email{aadlander@lowell.edu}

\author{Philip Massey}
\affiliation{Lowell Observatory, 1400 W Mars Hill Road, Flagstaff, AZ 86001, USA}
\affiliation{Department of Astronomy and Planetary Science, Northern Arizona University, Flagstaff, AZ, 86011-6010, USA}
%\email{phil.massey@lowell.edu}

\author{D. John Hillier}
\affiliation{Department of Physics and Astronomy \& Pittsburgh Particle Physics,
Astrophysics and Cosmology Center (PITT PACC), University of Pittsburgh,
3941 O'Hara Street, Pittsburgh, PA 15260, USA}
%\email{hillier@pitt.edu}

\author{Nidia Morrell}
\affiliation{Las Campanas Observatory, Carnegie Observatories, Casilla 601, La Serena, Chile}
%\email{nmorrell@carnegiescience.edu}

\begin{abstract}
We present a spectral analysis of four LMC WC-type Wolf-Rayet (WR) stars (BAT99-8, BAT99-9, BAT99-11, and BAT99-52) to shed light on two evolutionary questions surrounding massive stars.  The first is: are WO-type WR stars more oxygen enriched than the WC-type stars, indicating further chemical evolution, or are the strong high-excitation oxygen lines in the WO-type stars an indication of higher temperatures. This study will act as a baseline for answering the question of where WO-type stars fall in WR evolution.  Each star's spectrum, extending from 1100~\AA\ to 25000~\AA, was modeled using \cmfgen\ to determine the star's physical properties such as luminosity, mass-loss rate, and chemical abundances.  The oxygen abundance is a key evolutionary diagnostic, and with higher resolution data and an improved stellar atmosphere code, we found the oxygen abundance to be up to a factor of 5 lower than previous studies.  The second evolutionary question revolves around the formation of WR stars: Do they evolve by themselves or is a close companion star necessary for their formation?  Using our derived physical parameters, we compared our results to the Geneva single-star evolutionary models and the BPASS binary evolutionary models.  We found that both the Geneva solar metallicity models and BPASS LMC metallicity models are in agreement with the four WC-type stars, while the Geneva LMC metallicity models are not.  Therefore, these four WC4 stars could have been formed either via binary or single-star evolution.
\end{abstract}

%keywords{stars: individual (BAT99-8 [HD 32257, Brey 8, Sk -69$^\circ$ 42], BAT99-9 [HD 32125, Brey 7, Sk -66$^\circ$ 21], BAT99-11 [HD 32402, Brey 10, Sk -68$^\circ$ 15], BAT99-52 [HD 37026, Brey 43, Sk -67$^\circ$ 144]); stars: Wolf-Rayet, stars: evolution}

%% and \citet commands to identify citations. . 

\section{Introduction} \label{sec:intro}

Classical Wolf-Rayet stars (WRs) are the He-burning descendants of massive O-type stars.  Their hydrogen-rich outer layers have mostly been stripped away, revealing the evolved core of the star.  The spectra of WR stars are dominated by strong, broad emission lines formed in their optically thick stellar winds.  WRs are split into two subtypes, the WN-type and the WC-type, based on their optical spectrum.  The WN-type WRs show strong emission lines of helium and nitrogen (products of the earlier CNO-cycle H-burning), whereas the WC-type WRs show strong emission lines of carbon (products of He-burning via the triple-alpha process).   A subcategory of the WC-type WRs, or a possible third subtype, is the WO-type WRs (first discussed in \citealt{Barlow1982}), whose spectra are similar to those of the WCs, except they have stronger high-excitation oxygen lines (i.e., O\,{\sc vi}).  These strong oxygen lines could be a product of further evolution, indicating the WOs are more evolved than the WCs.  However, these lines could also suggest that the WOs have higher temperatures than the WCs, as that would result in the spectral oxygen lines strengthening as well.

The relation between the WC-type and WO-type WR stars can be determined by evaluating and comparing their physical parameters (luminosity, mass-loss rate, temperature, chemical abundances, etc.) using spectral modeling.  This paper introduces the baseline to this study, by modeling and evaluating four WCs in the Large Magellanic Cloud (LMC).  A follow-up study on a similar sample of LMC WO-type WRs is currently underway.

Already on its own, this study provides insight into the formation of WC-type WR stars.  Yet, the stripping mechanism that exposes the core of a WR is undetermined.   \cite{P1967} proposed that all WR stars have binary companions which were responsible for stripping off the outer layers of the star.  The ``Conti scenario'' \citep{Conti1976} argued that the mass-loss could be attained solely with stellar winds during earlier phases (i.e. O, Of) of evolution.

 The importance of binarity for the formation of WR stars remains unknown.  The frequency of short-period WR binaries in the LMC is roughly 30\% \citep{Foellmi2003-LMC}, similar to that in the Small Magellanic Cloud \citep{Foellmi2003-SMC}, the Milky Way \citep{vanderHucht2001}, and M31 and M33 \citep{ Neugent2014}.  However, the drawback of using the binary frequency to answer the formation mechanism is that we do not know if the other 70\% of WR stars were in binary systems that have since merged or if the binary companion is completely irrelevant in the stripping process.  \cite{Massey1981} compared the orbital separations of WR+O binaries and O-type binaries and found that mass transfer did not play an important role in WR evolution, but this study did not take into account mergers.  \cite{Sana2012} evaluated O-type stars for binarity and found that 70\% have a companion star, however, 1/3 of those will merge leaving their descendants to be single-stars.

In this study, we will be building off previous work analyzing WC stars by using significantly better observational data and a better analysis technique.  Past analyses (see e.g. \citealt{Hillier1999}, \citealt{Grafener1998}, \citealt{Crowther2002}, \citealt{2006ASPC..353..243B}, \citealt{Sander2012}) have shown that WCs are carbon-rich ($\sim$ 40\% carbon mass fraction) and have determined values for their other physical parameters.  Here, we will be using data obtained from the 6.5-m Magellan Baade telescope for the optical and near-Infrared wavelengths and data from the Hubble Space Telescope for the ultraviolet wavelengths.  Our modeling will utilize the stellar atmosphere code \cmfgen, which is constantly being improved upon (\citealt{Hillier1998}, \citealt{Hillier2003}, \citealt{Hillier2012}), and a tailored analysis, instead of a grid of models to give us unique physical parameters for each star and increased certainty in those values.  This is particularly important in establishing the baseline for a comparison to the WO stars we will analyze in the subsequent study.

Our sample consists of four WC4s (BAT99-8, BAT99-9, BAT99-11, and BAT99-52) in the LMC (listed in Table~\ref{tab:stars}).  A radial velocity study by \cite{Bartzakos2001}, found no signs of binarity for these four stars.  We chose stars in the LMC over the Milky Way, since the distance to the LMC is accurately known, and since LMC stars typically have much lower reddening than Milky Way WC stars.  A key result from our analyses, discussed separately in \cite{Hillier2021}, is that BAT99-9 contains nitrogen.

In the next section, we discuss the observations utilized in this project.  In Section 3, we discuss \cmfgen\ and the modeling process.  The physical parameters of these stars are given in Section 4.  The comparison to binary and single-star evolution models is in Section 5; our conclusions are provided in Section 6.  In the Appendix, we address four concerns with our stellar atmosphere code and discuss the atomic data used in \cmfgen.

\begin{deluxetable}{l l c c c c c c}
\tablecaption{\label{tab:stars} WC stars Observed}
\tablewidth{0pt}
\tablehead{
\colhead{ID}
&\colhead{Spectral Subtype}
&\colhead{$\alpha_{\rm 2000}$}
&\colhead{$\delta_{\rm 2000}$}
&\colhead{V(mag)}
&\colhead{M$_V$(mag)$^*$}
}
\startdata
BAT99-8 & WC4 & 04:56:02.89  & -69:27:21.5  & 14.23 & -4.7 \\
BAT99-9 & WC4 & 04:56:11.10  & -66:17:33.06  & 14.33 & -4.7 \\
BAT99-11 & WC4 & 04:57:24.08  & -68:23:57.1  & 12.69 & -6.0 \\
BAT99-52 & WC4 & 05:30:12.16  & -67:26:08.3  & 13.53 & -5.3 \\
\enddata
\tablecomments{$^*$M$_V$ has been corrected for extinction using the small reddening corrections determined in Section 4 and assuming a true distance modulus to the LMC of 18.5 (50 kpc). \\
All data except for $M_V$ comes from \cite{2018ApJ...863..181N} and references therein.}
\end{deluxetable}

\section{Observations and Reductions}

We obtained UV, optical, and NIR spectroscopic data for the four WC4 stars listed in Table~\ref{tab:stars}, giving us coverage from 1100~\AA\ through 25000~\AA. 

The UV observations, obtained with the Faint Object Spectrograph on the Hubble Space Telescope ({\it HST}), covered the wavelength region of 1140-3300~\AA\ at a resolving power $R=\lambda/\Delta \lambda$ of 1100-1600.  They were obtained using the G130H, G190H, and G270H gratings, with exposure times of 1500~s, 650~s, and 250~s respectively.  The observations were obtained on UT 1995 Nov 16 (data set Y2JE0403-7T), 1994 Sept 26 (data set Y2A10203-6T), 1996 April 15 (data set Y21E0105-8T), and 1995 Nov 21 (data set Y2JE0203-8T) for the stars BAT99-8, 9, 11, and 52, respectively.  These UV observations were obtained as part of GO-5460 (PI: Hillier).  The data were reduced using the standard reduction pipeline.

The optical data were collected using the Magellan Echellette  (MagE; see \citealt{MagE}) on the 6.5-m Magellan Baade Telescope at Las Campanas Observatory, covering the wavelength range 3150 to 9300~\AA.  The observations utilized the 1\farcs0 slit yielding a resolving power $R=4100$.  Each observation consisted of three exposures with the star moved along the slit in order to improve the signal-to-noise.  The slit was oriented to the parallactic angle.  The BAT99-8 observations were taken on UT 2016 February 21 at a 1\farcs0 seeing, an airmass of 1.4, and with 3$\times$600~s exposures.  BAT99-9 was taken on UT 2016 January 11 at a 0\farcs7 seeing, an 1.7 airmass, and with 3$\times$900~s exposures.  BAT99-11 was taken on UT 2016 February 21 at a 0\farcs89, an 1.4 airmass, and with 400~s exposures.  BAT99-52 was taken on 2015 November 26 at 1\farcs5 seeing, an airmass of 1.3, and with 3$\times$300~s exposures.  For each night of observing, some spectrophotometric standards were observed both at the beginning and end of the night.  A ThAr arc was used for the wavelength calibration.  For an in depth explanation of the reduction process, see \citet{2014ApJ...788...83M}.

The NIR spectra were taken using the Folded port InfraRed Echellette (FIRE) on the Magellan Baade Telescope, for a wavelength range 8300 to 25000~\AA.  The 0\farcs75 slit was used to obtain a spectral resolution of $R\sim5000$.  For the FIRE observations, four exposures are taken in a standard A-B-B-A pattern.   BAT99-8, BAT99-9, and BAT99-52 were all observed on UT 2016 February 20, with BAT99-11 being observed the following night. The four stars had an average airmass of 1.5 and an average seeing of 0\farcs9.  The telluric standards, HD~40750, HIP~23252, HD~46630, and HD~40925, were observed using four 30~s exposures to make the telluric correction and use as a flux calibration for BAT99-9, BAT99-8, BAT99-52, and BAT99-11, respectively.  These standards were observed directly after their corresponding WC star and at similar airmasses.   A ThAr lamp was used to make wavelength calibrations and data reduction was done using the {\sc idl} FIRE pipeline \citep{Simcoe2013}.

For each of the four WCs, we combined the UV, optical, and NIR fluxed spectra with the latter two scaled. Both the optical and the NIR observations will suffer slit losses that may be larger or smaller than the spectrophotometric standards used for the flux calibrations, while the {\it HST} UV fluxes should be accurate at the 1-2\% level. 
We therefore first scaled the optical spectra to the UV spectra by comparing the fluxes in the region of overlap, typically  3150-3300~\AA.   Similarly, the NIR spectra were then scaled to the rescaled optical in the region of overlap, typically 8300-9300~\AA.   We then combined the scaled spectra with the UV spectrum so that we had a single spectrum to work with.  We estimate that this correction process is good at the 10\% level or better, and is limited by the fact that the regions of overlap are always where the instruments are least sensitive and the calibration the poorest.  We emphasize, however, that this scaling and combining was done purely for convenience; in doing the fits, we rescaled as needed when fitting a particular region. These rescalings were typically $\pm$7\%, consistent with our expectations.

\section{Modeling}

Deriving the physical properties of WR stars is more complicated than generally true in stellar analysis work.  The spectra of most stars consist of absorption lines that are formed in a relatively thin, hydrostatic layer, where the geometry can be assumed to be plane-parallel.  Except for the hottest stars, and some unusual circumstances, local thermodynamic equilibrium (LTE) can be assumed.    The outer layers of such stars are composed of $\sim$73\% (by mass) of hydrogen and 25\% helium, with all other elements being a minor constituent by mass.  Even for massive OB stars where line formation is affected by mass-loss from stellar winds, the process is simple relative to that of WR stars, with the surface gravity being determined from the pressure broadening of H$\gamma$, the effective temperature from the strengths of He\,{\sc i} and He\,{\sc ii} lines \citep{Kudritzki1978}, and the mass-loss rates from emission components of H$\alpha$ and He\,{\sc ii} $\lambda 4686$, with the stellar wind terminal velocity $v_\infty$ (the velocity of the stellar wind far from the star where it is no longer accelerating) measured from the UV resonance lines (see, e.g., \citealt{Massey2013-bake-off}).  The luminosity then comes from the $V$-band photometry, after correction for reddening, distance, and after applying a bolometric correction based upon the temperature.  The Stefan-Boltzman equation can then be used to determine the stellar radius, and, using the surface gravity, the stellar mass.

By contrast, the spectra of WR stars consist of mostly emission lines, which are formed in an optically-thick  outflowing stellar wind. Mass-loss rates are typically an order of magnitude greater than that of even the most luminous O-type stars, and there is no such thing as a ``photosphere.''  Rather, high excitation lines are formed in the inner part of the wind, while lower excitation lines are formed in the outer part of the wind\footnote{The line-formation mechanisms have been extensively discussed by \cite{Hillier2015}.}   Further, the stellar  ``radius,'' ``surface gravity,'' and  ``effective temperature'' are no longer fundamental parameters since they depend on the properties of the stellar wind. One can still define the effective
temperature using the radius of the star at a Rosseland optical depth of 2/3, but this is not easily related to the effective temperature
defined by stellar evolutionary models. Alternatively, the
above quantities can be well defined using  a ``core radius'', $R_*$, that represents the inner boundary of the stellar atmosphere where the expansion velocity is negligible (see, e.g., \citealt{Hamann1987}, \citealt{Schmutz1989}). However, for many WR stars the ``core radius'' cannot be uniquely determined (\citealt{1991IAUS..143...59H}; \citealt{Najarro1997}; \citealt{Hamann2004}).

The stellar wind velocity law is taken to be of the form $$v(r)=v_\infty(1-R_*/r)^\beta,$$ which is often referred to as the ``Beta-law'' (adapted from \citealt{CAK1975}).  The value of $\beta$ determines how steeply the wind is accelerating, with larger values being less steep. In general, a value of $\beta=0.8$ seems appropriate for OB stars (although O super giants often have a $\beta>1.0$), while $\beta=1.0$ is taken for WR stars \citep{Hamann1987}.

For WC stars, \cite{Hillier1989} showed that the wind velocity undergoes a substantial increase at large radii ($r>10R_*$), and that thus a two-component velocity law results in somewhat better fits, with the outer region having a very shallow slope ($\beta=20$).    In practice, $v_{\infty1}$ in the inner region is found by fitting the line widths of the high-excitation lines (such as C\,{\sc iv}), while $v_{\infty2}$ in the outer region is found by fitting the line widths of the lower-excitation lines (such as C\,{\sc iii}).  The two-component velocity-law has been widely adopted by the massive star community, following the work of \cite{Grafener2005}.   Their hydrodynamic model showed that a two-component model was a good approximation.  More recently, \cite{Sander2015} used a method by which the velocity field was constantly updated by integrating the hydrodynamic equation for both inwards and outwards from the sonic point.

 The situation in terms of abundances is also complicated.  For many WN-type WRs a reasonable first approximation is that the surface composition is helium with little or no hydrogen, with most of the initial carbon and oxygen tied up in the form of nitrogen, in accord with our expectations of CNO-equilibrium abundances.  For most other species, standard abundances, can be adopted.

 For WC-type WRs the situation is far more complicated, since He-burning has produced carbon, giving it an abundance comparable to that of He.  The mass-fraction of carbon is dependent  on the degree of chemical evolution, and has to be determined concurrently with the other stellar parameters.  Furthermore, as carbon is built up in the core, helium and carbon will combine to produce oxygen, and potentially its abundance also has to be determined concurrently with the other stellar parameters.

\subsection{CMFGEN}
In order to model the stellar atmospheres of WC stars, a radiative transfer code that uses non local-thermodynamic-equilibrium (non-LTE) and line blanketing for extended, outflowing atmospheres is required.  \cmfgen\ (``Co-Moving Frame GENeral,'' \citealt{Hillier1998}) accomplishes this and was used to model the four WC stars.  The \cmfgen\ input parameters that we adjusted during our analyses are the luminosity, the terminal velocity of the wind and its shape, the mass-loss rate, and the chemical abundances of helium, carbon, and oxygen.

Determining the best values for each parameter was done using a tailored analysis.  This approach entails finding the best values by running model after model, altering the parameters and improving the fit until we have converged on a model with excellent agreement to the observed spectrum (as our group has done in the past, see e.g., \citealt{Hillier1989}, \citealt{Hillier1999}).  While this process takes longer than using a grid of models, it allows the model to be changed uniquely for each star to find the best values for each parameter.  It also allows us to evaluate more parameters than would be possible using a grid.

\cmfgen\ has also evolved throughout the years (\citealt{Hillier1998}, \citealt{Hillier2003}, \citealt{Hillier2012}) leaving a more accurate stellar atmosphere code than what has been used in previous studies of WR stars, by incorporating improved atomic data, using a two-component velocity law for WC stars, and in an improved treatment of clumping, to name just a few refinements.

\hspace{24pt}
\subsection{Fitting the Observed Spectrum}

Fitting the spectra of the WC-type stars is often a balancing act between all of the physical parameters.  This is due to the spectral lines being influenced by multiple parameters, e.g. luminosity, mass-loss rate, temperature, and chemical abundances. Further, most of the lines are blended, consisting of a combination of carbon, helium, or oxygen, such as all of the He\,{\sc ii} lines are blended with carbon lines.  \cmfgen\ may also contain errors in the atomic data which will influence the fit quality.  All of these things result in lines of the same species and ionization states not being equally well fit using a single set of parameters.  Thus, compromises are made during the process to obtain the best fit model.

In our WC spectra the continuum is masked by the rich emission line spectrum and this makes normalizing the spectrum futile.  Therefore, we did our fitting without normalization, utilizing fluxed spectra.  In order to allow for small differences in the fluxing from the UV, optical, and NIR, the models were relatively scaled to three regions where there are no emission lines to determine if the model fits the observed spectrum.  The UV, optical, and NIR model spectra were scaled using the relatively line-free regions from 1740-1780~\AA, 6000-6100~\AA, and 15100-15200~\AA, respectively.  When the scaling needed to be adjusted for large differences in the flux between the spectrum and the model, absolute scaling was done by adjusting the mass-loss rate, radius, and luminosity such that $\frac{\dot{M}^2}{R_*^3} = constant$ and $L \propto T_{eff}^4 R_*^2$.

The modeling process begins with finding reasonable values for the luminosity and the mass-loss rate.  Both are primarily determined by the continuum and all of the emission lines.  However, more explicitly, since the temperature and thus luminosity are tied to the distance at which lines form in the wind, the luminosity can be determined using the ionization ratios, e.g. C\,{\sc iv}/C\,{\sc iii}, O\,{\sc v}/O\,{\sc iv}/O\,{\sc iii}, and He\,{\sc ii}/He\,{\sc i}.  We found that the C\,{\sc iii} $\lambda$9710 line's flux was particularly useful in determining the luminosity.  Similarly, the flux of the He\,{\sc i} $\lambda$10830 is sensitive to the mass-loss rate of the star and can be used to further constrain its value.  The model is not sensitive to the given radius of the star (see Appendix A) and is thus only altered to adjust the absolute scaling.

It is well known that the winds of WR stars are clumped (e.g., \citealt{Owocki1988}; \citealt{Hillier1991b}, \citealt{Robert1994}; \citealt{Crowther2002}).  % {\red REFS}. 
Typically in \cmfgen\ we assume that there is no interclump medium, and describe the clumping using the volume-filling factor, ``f'' which has the form
\begin{equation}
f= f_\infty + (1-f_\infty)\exp(-V/V_{cl})
\end{equation}
where $f_\infty$ is the volume-filling factor at large radii, and $V_{cl}$ is an onset
onset velocity (see Section C in the Appendix). The above formula is not based on any physics -- it
is simply a convenient formulation, and other formulations could
easily be adopted.  For our
analyses, $f_\infty$ was typically  set to 0.05.  From the spectra, uniform clumping values of 0.2 and higher are found to be incompatible with  the red-wings seen on many emission lines (Appendix B and C). However, constraining $f_\infty$ is difficult, and it is also probable that our simple prescription does not adequately describe the run of $f(r)$ in the wind.

The velocity law terminal velocities were determined using line widths. The two-component velocity law used in \cmfgen\ requires us to use different lines  for each component.  In the inner region, where the temperature is the hottest, the terminal velocity component, $V_{\infty 1}$, is best determined using lines from moderate ionization species, such as C\,{\sc iv} and He\,{\sc ii}, and which account for the majority of the lines in the WCs' spectra.  After a general analysis of all of the high excitation lines in the spectrum, we used the width of C\,{\sc iv} $\lambda\lambda$5801,12 to refine the terminal velocity of the inner component.  For the outer region component, $V_{\infty 2}$, we used lines from ``low'' ionization species whose emission lines are formed in the outer region of the star's wind.  The lines we primarily used were C\,{\sc iii} $\lambda$2289, C\,{\sc iii} $\lambda$9710, and He\,{\sc i} $\lambda$10830.  The $\beta$ values were kept constant, with the outer component being set to $\beta = 20$ and the inner component being set to $\beta = 1.0$.

\begin{figure}[ht!]
\plotone{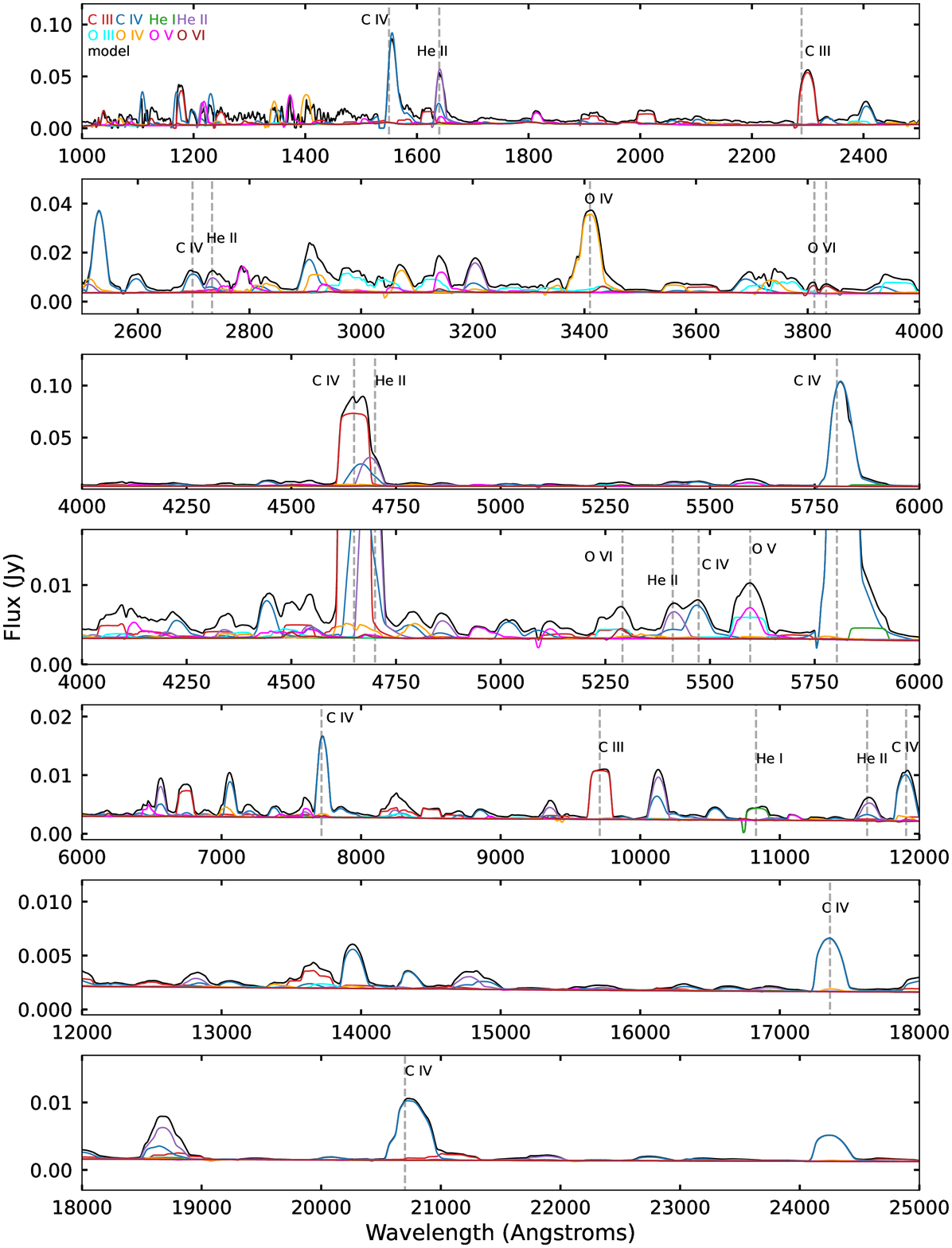}
\caption{Atomic species in BAT99-8's Model.  The main composition of the model's spectrum (black) features are He\,{\sc i} (green), He\,{\sc ii} (purple), C\,{\sc iii} (red), C\,{\sc iv} (blue), O\,{\sc iii} (cyan), O\,{\sc iv} (orange), O\,{\sc v} (magenta), and O\,{\sc vi} (brown).  Most of the lines in a WC star's spectrum are blended as can be seen with all of the atomic species contributions in the model separated out.
\label{Fig-species}}
\end{figure}

The chemical abundances of helium, carbon, and oxygen were found using corresponding emission line strengths and line ratios.  The line strengths of the emission lines are not only contributed to by the chemical abundance but also by luminosity (affects the C\,{\sc iii} to C\,{\sc iv} and He\,{\sc i} to He\,{\sc ii} line strengths), mass-loss (affected He\,{\sc i} $\lambda$10830) and temperature (affects the C\,{\sc iii} to C\,{\sc iv} and He\,{\sc i} to He\,{\sc ii} line strengths).  In Fig~\ref{Fig-species}, a WC model has been split into the different atomic species that contribute to the complexity of the spectrum.  Most lines in the WCs' spectra are blended lines with He contributing to some C lines and visa versa (for example, the He\,{\sc ii} $\lambda$4686 line has C\,{\sc iv} $\lambda$4683 and C\,{\sc iii} $\lambda$4673 contributing to it).  The atomic species that formed the \cmfgen\ model are discussed in Appendix E.

Therefore, once the line strengths are roughly correct from the four parameters, the relative abundances of carbon and helium are found using line ratios of C/He, C\,{\sc iv} $\lambda$2698/He\,{\sc ii} $\lambda$2733, C\,{\sc iv} $\lambda$5473/He\,{\sc ii} $\lambda$5411, and C\,{\sc iv} $\lambda$11905/He\,{\sc ii} $\lambda$11626.  In some cases (e.g. C\,{\sc iv} $\lambda$11905/He\,{\sc ii} $\lambda$11626) both lines tended to be systematically too strong; however, we assume that the ratio is still reflective of the relative abundance.  The oxygen abundance, on the other hand, was determined using the relative strength of the oxygen lines, primarily O\,{\sc vi} $\lambda$5291 and O\,{\sc v} $\lambda$5596.

For the reddening determination, we adopted a CMM law \citep{Cardelli1989} for the foreground component and a Mathis law for the LMC component \citep{Howarth1983}.  The Galactic foreground reddening law was set with values of R=3.1 and E(B-V)=0.08~mags \citep{Cardelli1989}.  We then fit the LMC component, finding E(B-V) values 0.03-0.08 mags.  We assumed a distance of 50~kpc for the LMC (\citealt{vandenBergh2000}, \citealt{Pietrzynski2019}).

The spectral fits to each star are shown in Figures~\ref{Fig-BAT99-8}-\ref{Fig-BAT99-52}. 
To emphasize both the complexity of the spectra, and
the futility of trying to normalize the spectra, we also show
the ``true" continuum of the model.

These model fits are representative of the WC stars, with the models matching even the low flux spectral features.  This is especially impressive as the models were adjusted to best fit the emission lines with the strongest fluxes.  The low flux emission features were analyzed last as a mere check for final agreement and parameters were only adjusted slightly for them.  The end result was a model that replicated almost every feature in the WC spectrum from their strength and shape.  The agreement of the small emission features shows the accuracy and capabilities of \cmfgen.

The models also reproduced the line widths of both the low and high excitation emission lines.  This was due to the two-component velocity law that enabled us to calibrate each component for the excitation lines separately.  

Each of the stars' corresponding model has features that agree particularly well.  In BAT99-9 and BAT99-11's models the oxygen lines, O\,{\sc vi} $\lambda$5291 and O\,{\sc v} $\lambda$5596, are in agreement with their spectrum line strength to within a factor of 1.05.  BAT99-9 and BAT99-52 have a fit to within a factor of 1.1 for C\,{\sc iii} $\lambda$2289, which was a difficult line to fit in BAT99-8 and BAT99-11 (factors of 1.3 off).  All of the models have a nice fit on the blend line of  C\,{\sc iv} $\lambda$4683 and He\,{\sc ii} $\lambda$4686 (factor of 1 to 1.2 different) and/or the C\,{\sc iv} $\lambda\lambda$5801,12 blend line (up to a factor of 1.2 different).

For the models, there were also some common problems.  For example, the O\,{\sc iv} $\lambda$3400 line is too high in all of the models, in some cases by a factor of 2.  The C\,{\sc iv} $\lambda$20700 line is consistently too high.  The C\,{\sc iii} lines, $\lambda$2289, and $\lambda$9710, had to be balanced as $\lambda$2289 was typically too high and $\lambda$9710 was typically too low.  This discrepancy can be attributed to a combination of errors in the atomic data  and slight errors in the models.

The O\,{\sc vi} $\lambda\lambda$3811,3834 doublet has additional issues in the models.  This doublet is an important feature that separates the WCs from the WOs.  The doublet is consistently too low in the models. This may be an affect of the clumping due to the ionization structure of the line.  The line is formed deeper in the stellar wind, meaning that the clumping of the models may be interfering with the photons making it to the surface.  Appendix D includes an in-depth discussion on the inability to fit the O\,{\sc vi} $\lambda\lambda$3811,3834 doublet.

\subsection{Chemical Abundance Uncertainties}
The chemical abundance uncertainties for the C/He ratio and O were determined by adjusting their values to find the limits where the fits would no longer be acceptable.  The C/He ratio uncertainty was determined using the He\,{\sc ii} and C\,{\sc iv} line pairs and referencing the He\,{\sc i} $\lambda$10830 and C\,{\sc iii} $\lambda$9710 lines.  This was found by changing the carbon and helium abundances of the best fit model for each star and leaving all other parameters fixed.  Once the line ratios were no longer representative of the line pairs (e.g. the line strengths became even or flipped), the uncertainty was called.  The oxygen was altered in a similar way, where the O\,{\sc vi} $\lambda$5291 and O\,{\sc v} $\lambda$5596 (with a significant O\,{\sc iii} contribution) lines were used to determine the realistic uncertainty.  The lower and upper oxygen uncertainties were established once the oxygen lines in the model were no longer fitting the oxygen lines.  These models showing the limits of acceptable values for the C/He abundance are shown in Figure~\ref{Fig-carbon-error} and for the oxygen abundance in Figure~\ref{Fig-oxygen-error}.  The uncertainty values for the chemical abundances are listed in Table~\ref{tab:PP}.

The oxygen abundance was also given a larger ``generous'' uncertainty to check the limits of oxygen in one of the stars (BAT99-8).  The lower limit was determined by fitting the O\,{\sc iv} $\lambda$3400 line, as all of the stars' best fit models were too high for this line.  However, by reducing the oxygen abundance to fit the O\,{\sc iv} $\lambda$3400 line, all other oxygen lines in the model became too low to be reasonable fits to the spectrum.  The upper limit for the ``generous'' oxygen uncertainty was found by making all of the oxygen lines high in the model compared to the spectrum.  This is also unreasonable for the actual oxygen abundance for the WC star.  Therefore, the generous uncertainly describes the possible range of oxygen, but not a reasonable range of oxygen for the star.  A model with an oxygen abundance at either of these bounds would not be a good fit to the spectrum and instead is used as a check on what the extent of oxygen could be in the star.  The resulting generous uncertainty for BAT99-8 was $0.077^{+0.13}_{-0.037}$.

\subsection{Hidden Companions}
	
The four stars analyzed here were selected as none show signs of having a binary companion.  None show either the radial velocity variations \citep{Bartzakos2001} or any spectroscopic sign of a companion \citep{Bartzakos2001,2018ApJ...863..181N}. Nevertheless, as the late Virpi Niemela would remind us, one can only prove that a particular star {\it is} a binary, never that it is single.  The orbital inclination may be unfavorable for radial velocity variations, or the companion might be too faint to detect spectroscopically.  We were therefore interested in understanding what sort of companion could have escaped our notice.

To address this, we utilized the TLUSTY BSTAR2006 \citep{2007ApJS..169...83L}\footnote{\href{http://tlusty.oca.eu/Tlusty2002/tlusty-frames-BS06.html}{http://tlusty.oca.eu/Tlusty2002/tlusty-frames-BS06.html}}
and Phoenix \citep{2013A&A...553A...6H}\footnote{\href{http://phoenix.astro.physik.uni-goettingen.de}{http://phoenix.astro.physik.uni-goettingen.de}} models.
We considered the possibility that our spectra were actually blends  of WC4 stars plus a main-sequence companion star ranging from B0~V to A0~V.  For this experiment we considered companion stars of spectral types B0~V ($M_V=-4.0$, $T_{\rm eff}=$28,000~K, 18$M_\odot$), 
B2~V ($M_V=-2.5$, $T_{\rm eff}=$18,000~K, 10$M_\odot$), B5~V ($M_V=-1.2$, $T_{\rm eff}=$14,000~K, 6$M_\odot$), B8~V ($M_V=-0.3$, $T_{\rm eff}=$11,000~K, 4$M_\odot$), and A0~V ($M_V=+0.7$, $T_{\rm eff}=$10,000~K, 3$M_\odot$), with the values drawn from  \citet{Allen} and \citet{HumphreysMcElroy}. The corresponding $\log g$=4.0 models at LMC-metallicity
were rotationally broadened to 75 km s$^{-1}$, and their fluxes scaled to the expected $M_V$.  The model spectra were then reddened using a CMM law with an $E(B-V)=0.13$, typical of what we find for our SED fitting  (Section 5) and the LMC in general \citep{LGGSII}.  We then subtracted the reddened models from our spectra and examined the resulting spectra. We found that the observations could easily hide an A0~V or a B5~V star, but that a companion of spectral type B2 V or earlier would distort the even-N Pickering He\,{\sc ii} lines due to the strong hydrogen lines present in the B stars.  Thus, we would be blind to any stellar companion with a mass of $\sim$8M$_\odot$ or lower.

We will also note that a close compact companion (neutron star or black hole) would also go undetected if the orbital inclination was seen pole-on.  However, a Vizier search does not find any of these stars associated with X-ray sources, suggesting that none of these stars has a close compact companion.  It is possible that the stars could be in a wide orbit binary, in which case the X-ray source may not be detected \citep{Eldridge2020}.  Another option is that the X-ray emission from the companion is weak; this could occur if the companion is a black hole without an accretion disk \citep{Schootemeijer2018}.

\section{Physical Parameters}

\begin{deluxetable}{l l c c c c c c c c c c c c c} 
\tablecaption{\label{tab:PP} Physical Parameters}
\tablewidth{0pt}
\tablehead{
\colhead{ID}
&\colhead{log $\dot{M}$} %Mass-Loss Rate ($M_\odot/yr$)
&\colhead{log(L/$L_\odot$)}
&\colhead{R$_{2/3}$ $^1$}
&\colhead{$T_{\rm eff}$ $^1$}
&\colhead{$f$}
&\colhead{$\log \dot{M}/\sqrt{f}$}
&\colhead{X(He)}
&\colhead{X(C)}
&\colhead{X(C)/X(He)}
&\colhead{X(O)}
&\colhead{X(Ne)}
&\colhead{E$_{B-V}^{LMC}$ $^2$}
&\colhead{$V_{\infty,1}, V_{\infty,2}$ $^3$} 
&\colhead{Ref$^4$} \\
& & & R$_\odot$ & $kK$ & & & & & & & & mags & km s$^{-1}$ &
}
\rotate
\tabletypesize{\scriptsize}
\startdata
BAT99-8  & -4.84 & 5.48 & 2.4 & 87 & 0.05 & -4.2 & 0.50 & 0.41 & 0.83$^{+0.25}_{-0.08}$ & 0.077$^{+0.063}_{-0.017}$ & 0.011 & 0.05 &  1600, 2600\\
 & \it{-4.06} & \it{5.13} & \it{6.46} & \it{43.6} & \it{-} & & \it{-} & \it{0.4} & & \it{0.3} &  \it{-} &  \it{0.06} &  \it{2300} & \it{G98}  \\
 & \it{-4.9} & \it{5.42} & \it{3.4} & \it{71} & \it{0.1} &  & \it{0.45} & \it{0.47} & & \it{0.08} &  \it{-} &  \it{-} &  \it{2300} & \it{C02} \\
  BAT99-9 & -4.85 & 5.48 & 2.6 & 84 & 0.05 & -4.2 & 0.66 & 0.29 & 0.43$^{+0.13}_{-0.04}$& 0.041$^{+0.037}_{-0.008}$&  0.011 & 0.08 & 1600, 2300\\
& \it{-4.16} & \it{5.29} & \it{5.94} & \it{49.8} & \it{-} & & \it{-} & \it{0.4} & & \it{0.2} &  \it{-} &  \it{0.08} &  \it{2300} & \it{G98}  \\
& \it{-4.8} & \it{5.44} & \it{3.2} & \it{74} & \it{0.1} &  & \it{0.65} & \it{0.25} & & \it{0.10} & \it{-} & \it{-} & \it{2500} & \it{C02}  \\
BAT99-11 & -4.42 & 5.86 & 5.5 & 72 & 0.05 & -3.8 & 0.72 & 0.25 & 0.34$^{+0.11}_{-0.03}$ & 0.025$^{+0.023}_{-0.0050}$ & 0.011 & 0.06 & 2100, 3000\\
 & \it{-3.80} & \it{5.62} & \it{10.4} & \it{45.6} & \it{-} & & \it{-} & \it{0.5} & & \it{0.2} &  \it{-} &  \it{0.05} &  \it{2800} & \it{G98} \\
 & \it{-4.5} & \it{5.70} & \it{5.2} & \it{67} & \it{0.1} & & \it{0.66} & \it{0.28} & & \it{0.05} &  \it{-} &  \it{-} &  \it{3000} & \it{C02} \\
BAT99-52 & -4.70 & 5.65 & 3.3 & 83 & 0.05 & -4.0 & 0.47 & 0.42 & 0.90$^{+0.27}_{-0.09}$ & 0.094$^{+0.076}_{-0.017}$ & 0.011 & 0.03 & 1800, 2600\\
  & \it{-3.91} & \it{5.26}& \it{8.03} & \it{42.1} & \it{-} & & \it{-} & \it{0.4} & & \it{0.3} & \it{-} & \it{0.01} & \it{2600} & \it{G98}\\
& \it{-4.5} & \it{5.65} & \it{4.8} & \it{68} & \it{0.1} &  & \it{0.46} & \it{0.44} & & \it{0.09} & \it{-} & \it{-} & \it{2900} & \it{C02}  \\
\enddata
\tablecomments{
$^1$ R$_{2/3}$ and $T_{\rm eff}$ are defined at a Rosseland optical depth of 2/3.\\
$^2$ Determined using a foreground reddening component E(B-V)=0.08 plus the LMC component listed here. \\
$^3$  We used a two-component velocity law, while the previous studies used a one-component law.\\
$^4$ References: G98 - \cite{Grafener1998}; C02 -\cite{Crowther2002}
}
\end{deluxetable}

The physical parameters determined by our fits are summarized in Table~\ref{tab:PP}, along with corresponding values from \cite{Grafener1998} and \cite{Crowther2002}.  The four WC stars have mass-loss rates that vary by a factor of 2.7.  Similarly, the luminosity of the stars differs by a factor of 2.4.  The chemical abundances are closer in value, with helium changing by a factor of 1.53 between the stars, carbon differing by a factor of 1.68, and oxygen being a factor of 2 different. 

While evaluating the four WC stars, we discovered three emission features in BAT99-9's spectrum that were not in the other WC stars' spectra, specifically, N\,{\sc v} $\lambda$1240, N\,{\sc iv} $\lambda$1718, and N\,{\sc iv} $\lambda$3480.  These lines were fit by introducing nitrogen into the model (0.00095 by mass).  This is the first discovery of nitrogen in a WC star, and the implications are discussed further in \cite{Hillier2021}.

\subsection{Comparison with Previous Work}

The four WC-type stars analyzed here were also modeled by \cite{Grafener1998} and \cite{Crowther2002}.  \cite{Grafener1998} used  the Potsdam Wolf-Rayet Models (PoWR, \citealt{Grafener2002}) and \cite{Crowther2002} utilized \cmfgen\ \citep{Hillier1998}.  However, the models utilized in \cite{Grafener1998} were not fully blanketed; i.e., the models did not account for blanketing by the iron group elements.

In terms of the helium abundances, we find similar values to those found by \cite{Crowther2002} with all differences $<$ 10\%.

In terms of the carbon abundances, we find similar values to \cite{Crowther2002}, with an agreement of 15\% or lower, but with larger differences with \cite{Grafener1998}.  BAT99-8 and BAT99-52 have carbon abundances that have an agreement within 5\% to \cite{Grafener1998}, but for BAT99-11, \cite{Grafener1998} has an abundance that is a factor of 2 larger.  The carbon abundance is in agreement with the limit proposed by \cite{Higgins2021}, in which a WC star can only ever reach a carbon mass fraction of 50\% regardless of the amount of stripping taking place (shown in Figure 5 of their text).

The biggest difference between these previous studies and ours is the oxygen abundance.  Our values are similar to \cite{Crowther2002} for two of the stars (BAT99-8 and 52), but half as large for BAT99-9 and 11.  Our values are a factor of 3 to 5 smaller than those of \cite{Grafener1998}.  The oxygen abundance is one of the most important values in our study due to the main difference between the WCs and WOs being an O\,{\sc vi} line.

The difference between these two sources and our results may be due to the improvements in our stellar atmosphere code and atomic data over the last two decades or our exceptional spectra data set.  

Finally, we compare our results with the general findings of
\citet{Sander2012}, who analyzed a sample of Galactic WC stars,
ranging in spectral subtype from WC4 to WC9\footnote{This landmark
paper was the first to use fully blanketed models to analyze a set
of WCs, but did suffer from several unavoidable problems. One
difficulty with such Galactic studies, particularly in the pre-Gaia
era, was the uncertainties in distance and hence the derived
luminosities. Furthermore, the typically high Galactic reddening
precluded UV observations of many of their targets.  Finally, their
optical data relied upon the \citet{TorresMassey} spectrophotometry,
which, although the best achievable in the pre-CCD era, suffered
from linearity issues for many of the stronger lines, as we have
found in comparing our modern data to theirs.} There are three WC4
stars in their sample; we find luminosities are considerably higher
($\log L/L_\odot$=5.5-5.9) than those in their sample ($\log
L/L_\odot=5.1-5.2$) even when corrected for revised distances from
Gaia \citep{2019A&A...621A..92S}.  Our mass-loss rates range from $\log
\dot{M}/\sqrt{f}$=-3.8 to -4.2; theirs are very similar, ranging
from -4.1 to -4.3.  They did not actually adjust their fits for
abundances, but rather adopted a carbon mass fraction of 40\%, 
55\% for helium, and 5\% for oxygen.  This can be compared
to our derived values in Table 2, where the mass fraction of carbon
ranges from 25\% to 42\%, that of helium ranges from 47\% to 72\%,
and that of oxygen range from 2.5\% to 9.4\%.  In \cite{Sander2012}'s Appendix, WR52 and WR144 were modeled using a higher oxygen mass fraction of 15\% with a carbon mass fraction of 50\% and a helium mass fraction of 35\%.  Similar to our models, even with a higher oxygen abundance the O\,{\sc iv} $\lambda\lambda$3811,3834 doublet was still not able to be reproduced (remaining too low in the models).   We expect that as
nuclear burning proceeds that the carbon and oxygen abundances will
increase at the expense of helium, and that shall find in the next
section that our abundance determinations serve as a powerful tool
for diagnosing the evolutionary status of these stars.

\section{Comparison to Evolutionary Models}

An important part of our study is to find out what we can learn about the evolution of these stars by comparing their physical properties to those predicted by stellar evolutionary models.  For this task, we used the ``Binary Population and Spectral Synthesis'' (BPASS) models \citep{Eldridge2017} and the Geneva single-star evolutionary models \citep{Ekstrom2012} to determine if our results are in agreement with either type of evolution. For BPASS, the LMC metallicity ($Z$=0.006) binary models were used; while the BPASS models contain single star evolutionary models as well, these were only used as a check and not in our evaluation.  We used the Geneva models that included rotation computed for metallicities of $Z$=0.006 (P. Eggenberger, 2021, in prep.) and $Z$=0.014 \citep{Ekstrom2012}.  Although the former is closer to the metallicity of the LMC ($Z$=0.007), the higher metallicity models also proved very useful for comparison, as they have higher mass-loss rates during the main-sequence phase.  This allows us to evaluate evolutionary tracks with a wider variety of mass-loss rates, in addition to accounting for uncertainties in the mass-loss evolutionary models' calculations.  

For the comparison to the WC star sample, we only used the evolutionary models that had a WC phase according to the limits used by \cite{Georgy2012}, X$_{\text{H}}$ $<$ 0.3 and log(T$_{\text{eff}}$) $>$ 4.0, and the standard requirement for the WC-type stars that C$>$N.  Binary stripping can produce hot stars with the expected surface chemical compositions as WRs, but whose luminosities are too low to develop the optically-thick stellar winds that are needed for a star to be identified as a WR (e.g. \citealt{2020MNRAS.491.4406S}, \citealt{Shenar2020}).  Thus, typically a luminosity limit of $\log L/L_\odot > 4.9$ is further proposed.  This is an issue for WN-like stripped binaries, but not for WCs, in that none of the BPASS models that met the other criteria had luminosities less than $\log L/L_\odot$=5.2.  The authors made use of a python package called \textit{hoki} \citep{hoki2020} to aid in the process of reading the BPASS models.

\subsection{Chemical Abundances}

In Fig.~\ref{Fig-He-C-O}, the WC stars' mass fractions of helium, carbon, and oxygen are compared to those of the evolutionary models.  As a WC star evolves, we expect that the carbon will increase and the helium will decrease as a result of the triple-alpha process.  As the carbon abundance increases, oxygen will be created as helium and carbon combine.  This is shown in the evolutionary tracks of the Geneva single-star and BPASS binary models.  The measured chemical abundances for the four WC stars fall directly on both the BPASS and Geneva evolutionary tracks.  Moreover, the chemical abundances of BAT99-8 and BAT99-52 indicate that they are more chemically evolved than BAT99-9 and BAT99-11.

In Fig.~\ref{Fig-chemcial_abundance}, we compare the measured surface abundance ratios with those predicted from the models.  Except for the 60 $M_\odot$ Z=0.006 model, the Geneva models predict a tight relationship between the O/He and C/He ratios, as would be expected for single-star evolution.  Our results fall along this sequence.  By contrast, the BPASS binary models show a large range of O/He values for a given C/He ratio.

The BPASS models, while spanning a larger range of O/He values, do have some that fall near our observed values.  The span of BPASS models could be due to binary interactions extending or changing the time of the helium burning to produce more oxygen in those stars.  Total, the BPASS models have around 3000 models that produce WC stars based on our cutoffs.  About a third (1133) of these models are in the region populated by the four WC stars studied here.  In this smaller region half the models have a companion star that has merged or is no longer visible (fainter than $M_V$ of -1 mag).  We can split this region up further by looking at WC masses that are greater and smaller than 21~$M_\odot$.  The models with primary stars less massive than 21~$M_\odot$ are the models closest to the four WC stars analysed here (BAT99-8, BAT99-9, BAT99-11, and BAT99-52), totalling 282 models with 185 models having faint companions.

\subsection{Luminosity}

 In Fig~\ref{Fig-L-CHe-OHe}, we compare our values for the luminosities and surface abundances with the evolutionary models. The luminosities of the Z=0.006 Geneva models do not extend as low as what we have for the observed stars.  However, there is good agreement with the higher metallicity Z=0.014 Geneva models.  Although BAT99-8 and -9 are of lower luminosities than the lowest mass Geneva model that still produces a WC star (40 $M_\odot$ solar metallicity model), this is likely a limitation of the large gaps between mass tracks, i.e., the models displayed here differ by at least 20 $M_\odot$. It is possible that a 35 $M_\odot$ model would produce a WC star with a lower luminosity needed for BAT99-8 and BAT99-9.

The BPASS models produce WC stars with luminosities extending down to those of the WC stars.  The region of BPASS models near the stars in the chemical abundance parameter space also span a region between $\log(L/L_\odot) = 5.3 - 6.4$ dex. This indicates that those models could produce the four WC stars studied here.

There are no WC-type stars formed from either the Geneva or the BPASS models below log(L/L$_\odot$) of 5.2 dex.  In the LMC, we have not observed any WC-type stars that have lower luminosities, therefore, this lower limit in the evolutionary models is in agreement with reality.

The luminosity has also been compared to the current WC masses of the Geneva and BPASS evolutionary models as seen in Fig~\ref{Fig-logL-m}.  The evolutionary tracks of the Geneva models and BPASS models have the same luminosity-mass relation.  The selected BPASS models in the region similar to our chemical abundance parameter space are in agreement with the solar metallicity Geneva models.  This suggests that, the physical parameters of WC stars are independent of how they formed.

\subsection{Lifetimes}
The chemical abundance ratios vs.\ time before the star goes supernova is shown in Fig~\ref{Fig-t-CHe-OHe}.  According to both the Geneva and BPASS models, the chemical abundances of the WC stars can be achieved at one-third of the WCs' lifetime for BAT99-9 and 11 and at about two-thirds of the WCs' lifetime for BAT99-8 and 52.  The time (before the star goes supernova) at which the stars' C/He and O/He agree with the Geneva evolutionary models occur at roughly the same time: 0.13 Myr, 0.14 Myr, 0.17 Myr, 0.18 Myr for BAT99-52, BAT99-8, BAT99-9, and BAT99-11 respectively.  This is also true for the BPASS models which happen within a few hundredths of a Myr from the time the Geneva models match.  The time at which these chemical abundances occur in the models indicates that both sets of models can predict the observed chemical abundances for a WC star.  The length of time for each of these models to have the observed chemical abundances is also reasonable.

\subsection{Mass-Loss Rates}
The mass-loss rates assumed by the evolutionary BPASS and Geneva models are consistent with those determined for the four WCs given our clumping factor, shown in Fig~\ref{Fig-massloss}.  When the mass-loss rate is compared with the luminosity, the stars are in agreement with the Geneva solar metallicity models, except the 85 $M_\odot$ model.  The BPASS models with similar chemical abundances to the four WC stars have a wide range of mass-loss rates, but most fall in the region with mass-loss rates and luminosities comparable to the four WC stars.

The C/He ratio compared to the mass-loss rate shows excellent agreement for BAT99-11 which falls between the Geneva solar metallicity models.  The other WC stars have mass-loss rates that are lower than what the Geneva models have for their C/He abundances.  However, a slightly lower mass Geneva model would take into account BAT99-52 position and possibly BAT99-8 and BAT99-9.  The BPASS models are in excellent agreement with the WC stars.  While, the Geneva models do not correspond with two of the WC stars, this is not enough to rule out single-star formation for WC stars.

\section{Summary and Conclusions}

In this paper, we have derived physical parameters for four apparently single WC stars in the LMC: BAT99-8, BAT99-9, BAT99-11, and BAT99-52.  We have compared these properties to those predicted both by the Geneva single-star models and the BPASS binary models. 

We find a significant lower oxygen abundance than found by previous studies of these four stars (\citealt{Grafener1998}, \citealt{Crowther2002}).  The explanation for this difference may be that we used higher resolution data and an improved stellar atmosphere code.  The oxygen abundance is significant for determining how these stars evolved and the relation between WC and WO stars.

Both sets of Geneva models do a good job of matching the surface abundances of the WC stars. However, the lower metallicity models, $Z=0.006$, do not predict WC stars with as low luminosities as what we observe, despite being a better match to the metallicity of the LMC.  This is likely due to the difference in the adopted mass-loss rates during the evolution of these stars.  One explanation for this could be that the Geneva models assume a stable mass-loss rate, while in actuality these stars may also experience mass-loss through eruptions.  These eruptions would mean that the Wolf-Rayet stars are undergoing more mass-loss than the Geneva models account for.  We should also keep in mind that although the oxygen abundance of the gas in the LMC is well determined, the overall metallicity is not set in stone, as the $Z=0.007$ value assumes scaling of solar abundances, which themselves have been a matter of debate (\citealt{2005ASPC..336...25A}, \citealt{2009ARA&A..47..481A}).

The BPASS models span an expansive range in chemical abundances, luminosities, and star masses.  However, only a third of the BPASS models that produce a WC star have similar chemical abundances to the observed WC stars.  The BPASS models in this region have reasonable luminosities, timescales, mass-loss rates, etc. for the WC-type stars and half of them no longer have a companion star or have a faint companion.

From our comparisons of the WC stars' physical parameters to both the Geneva single-star and the BPASS binary evolutionary models, we found that these stars are in a parameter space compatible with models from both types of evolution.  In fact, it may not matter how WC-type stars lost their mass as both types of evolutionary models have the same luminosity-mass relation.

The next step in this study is to model our sample of WO-type stars and compare their physical parameters with those found here for the WC-type stars.  This upcoming study will allow us to assess if the WO-type WR stars are more chemically evolved than WC-type stars or are at a similar stage of evolution.

\acknowledgments

 Northern Arizona University and Lowell Observatory sit at the base of mountains sacred to tribes throughout the region. We honor their past, present, and future generations, who have lived here for millennia and will forever call this place home.  The optical and NIR spectra used in this study were obtained at Las Campanas Observatory (LCO), and we are grateful to the continued support of the Carnegie and Arizona Time Allocation Committees for our work.  We are also grateful to the excellent support and technical assistance we also receive at LCO.  This work was partially supported through the NASA/ADAP grant 80NSSC18K0729, through the National Science Foundation grant AST-1612874, and through the STScI grant HST-GO-13781.  Computer resources were also supported through a Slipher Society award. EA would like to acknowledge the gracious support of her Lowell Pre-doctoral Fellowship by the BF Foundation.  We also thank Kathryn Neugent for help both with some of the observing proposals and for taking some of the data used here. The authors are grateful to Jan Eldridge and Andreas Sander for useful suggestions on a previous version of this manuscript.  We thank our anonymous reviewer for the helpful feedback on our paper.

\clearpage

\begin{figure}[ht!]
\plotone{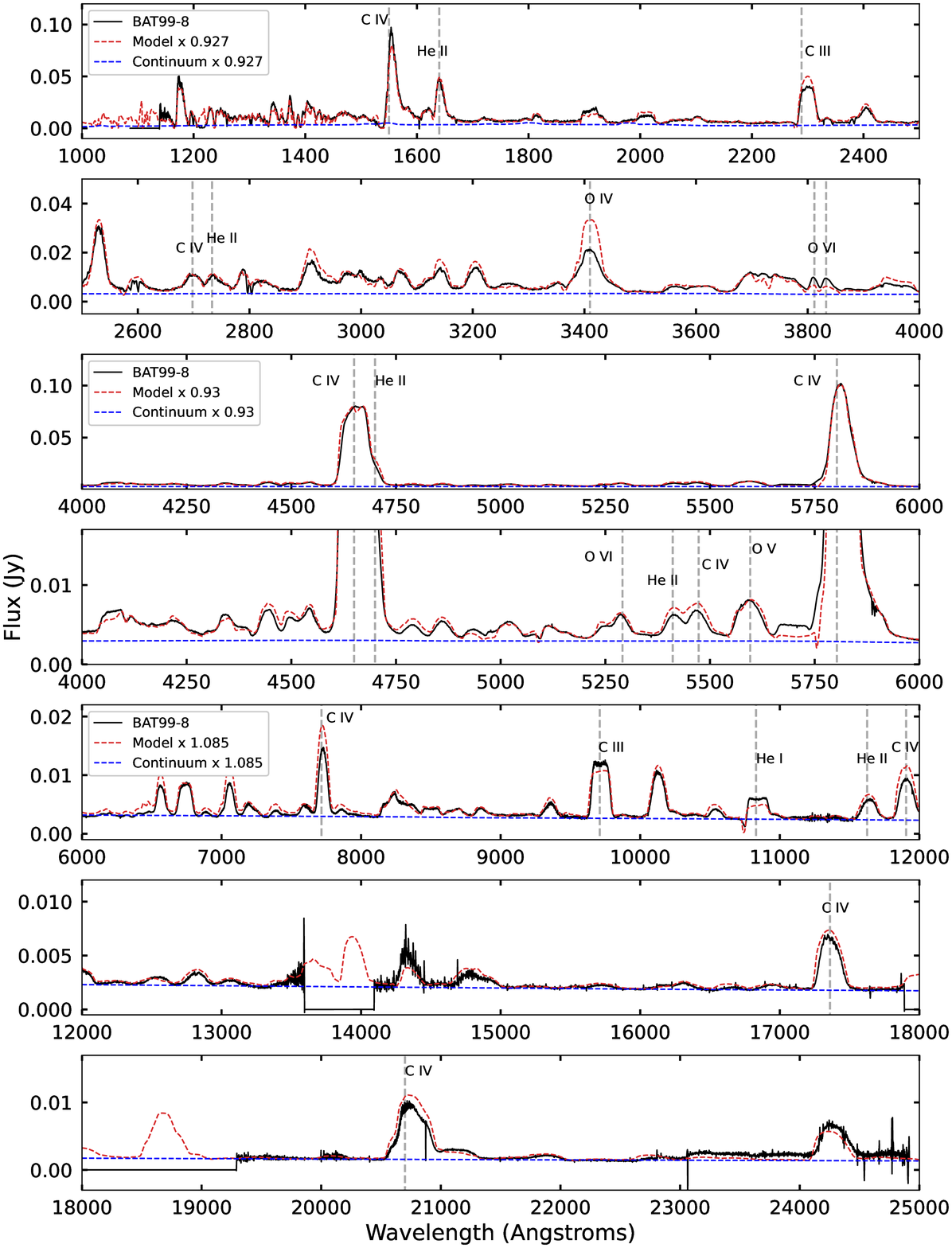}
\caption{The observed spectrum of BAT99-8 (black) with the best fit model (red dashed) and the model's continuum (blue dashed).  The model was scaled to the spectrum's continuum.  The labeled emission lines are those primarily used for modeling the spectrum.  The current disagreements in the model being too high for O\,{\sc iv} $\lambda$3400 and too low for O\,{\sc vi} $\lambda\lambda$3811,3834  may be due to atomic data issues in \cmfgen.
\label{Fig-BAT99-8}}
\end{figure}

\begin{figure}[ht!]
\plotone{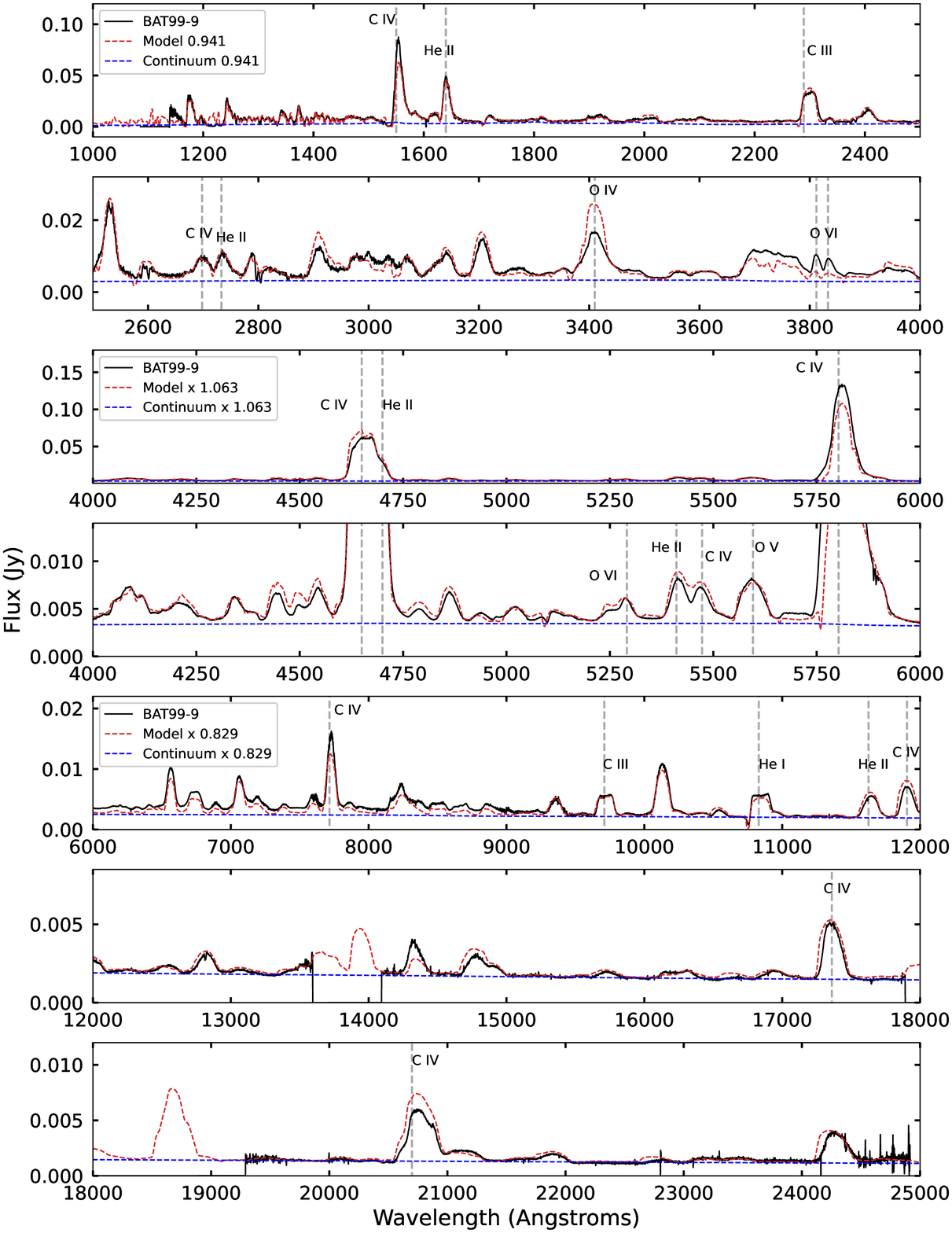}
\caption{The observed spectrum of BAT99-9 (black) with the best fit model (red dashed) and the model's continuum (blue dashed).  The model was scaled to the spectrum's continuum.  The labeled emission lines are those primarily used for modeling the spectrum.  The model's physical parameters are compromised to get the best fit model; therefore, there are some features in the spectrum that are not fit as well as others.
\label{Fig-BAT99-9}}
\end{figure}

\begin{figure}[ht!]
\plotone{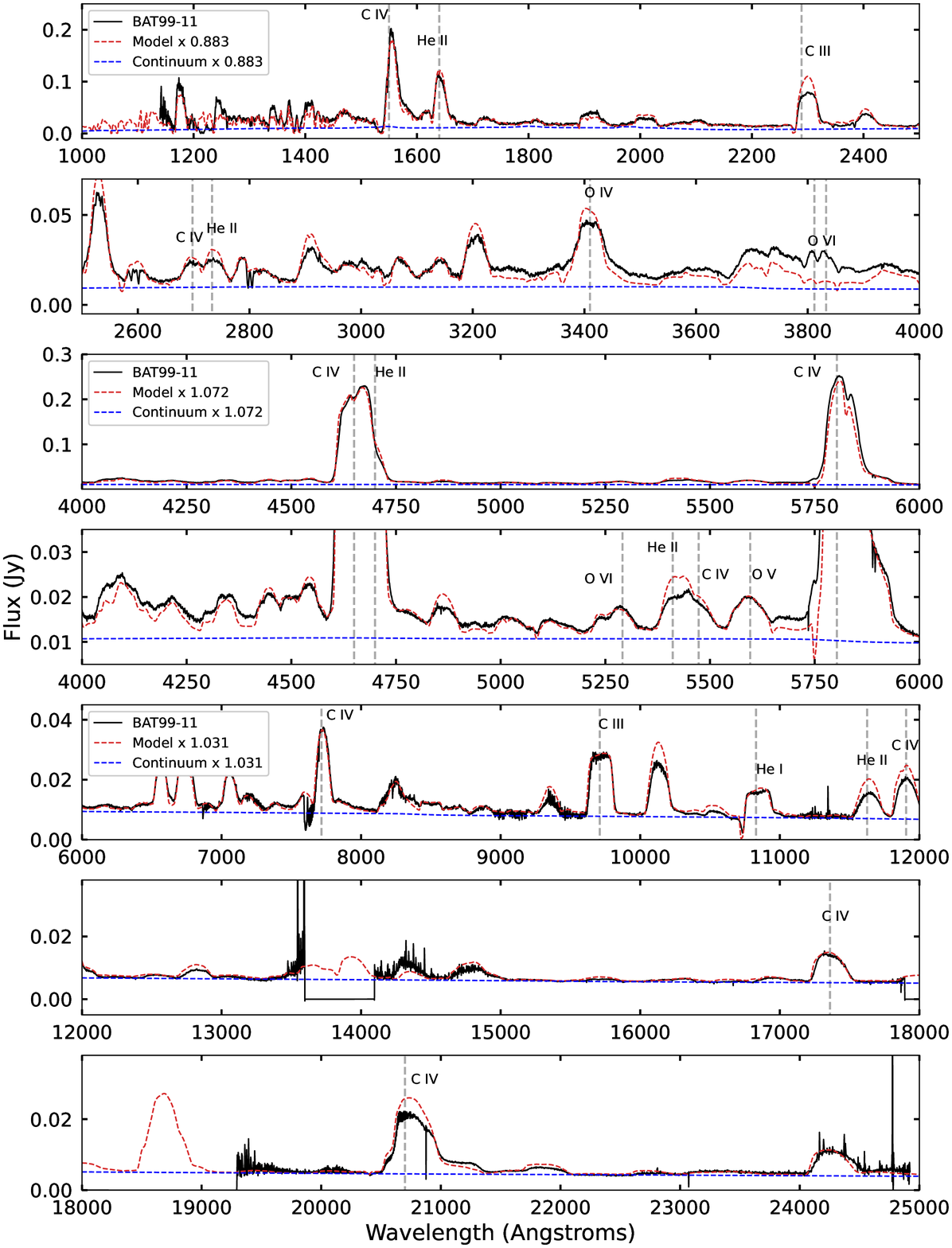}
\caption{The observed spectrum of BAT99-11 (black) with the best fit model (red dashed) and the model's continuum (blue dashed).  The model was scaled to the spectrum's continuum.  The labeled emission lines are those primarily used for modeling the spectrum.  The model's physical parameters are compromised to get the best fit model; therefore, there are some features in the spectrum that are not fit as well as others.
\label{Fig-BAT99-11}}
\end{figure}

\begin{figure}[ht!]
\plotone{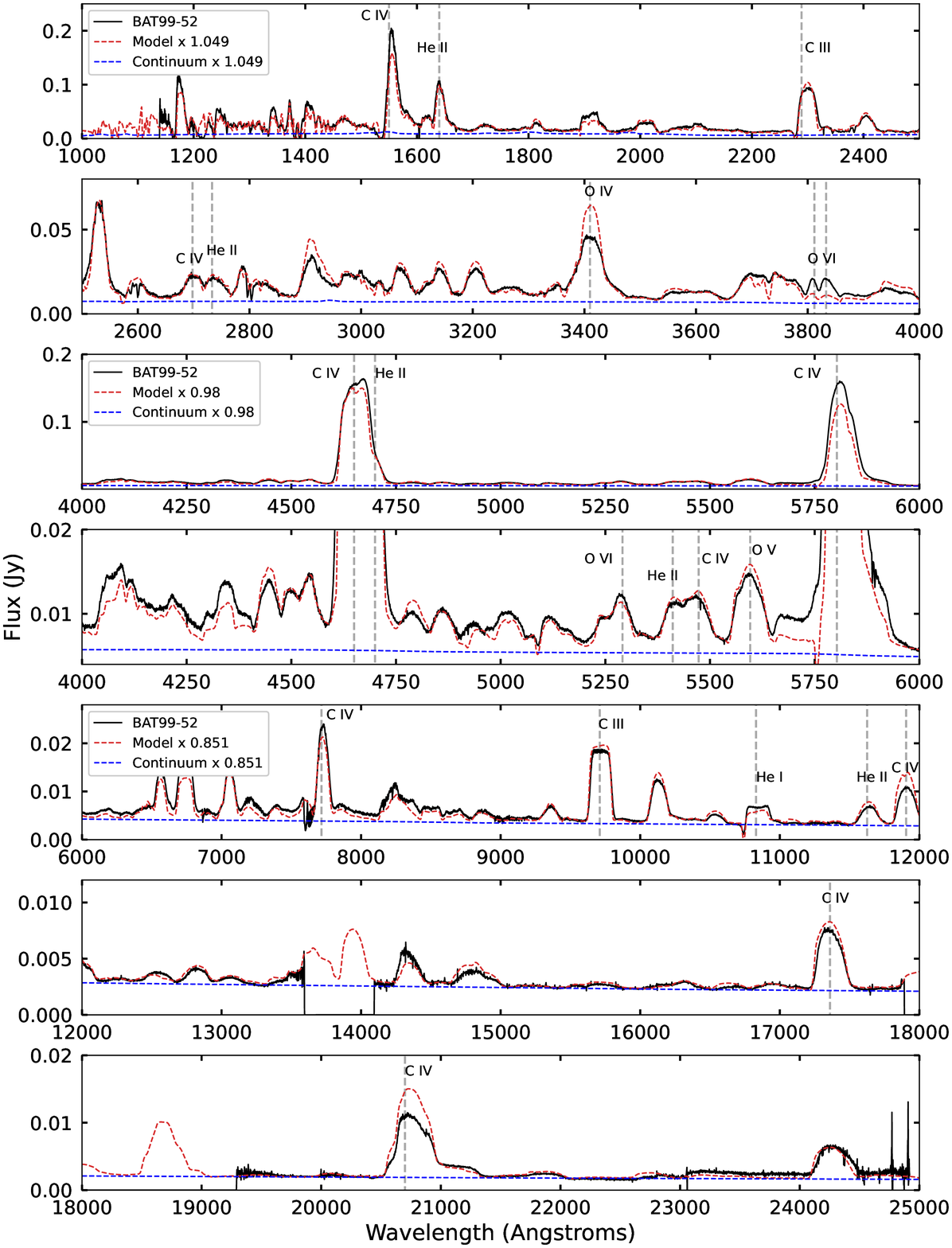}
\caption{The observed spectrum of BAT99-52 (black) with the best fit model (red dashed) and the model's continuum (blue dashed).  The model was scaled to the spectrum's continuum.  The labeled emission lines are those primarily used for modeling the spectrum.  The model's physical parameters are compromised to get the best fit model; therefore, there are some features in the spectrum that are not fit as well as others.
\label{Fig-BAT99-52}}
\end{figure}

\begin{figure}[ht!]
\plotone{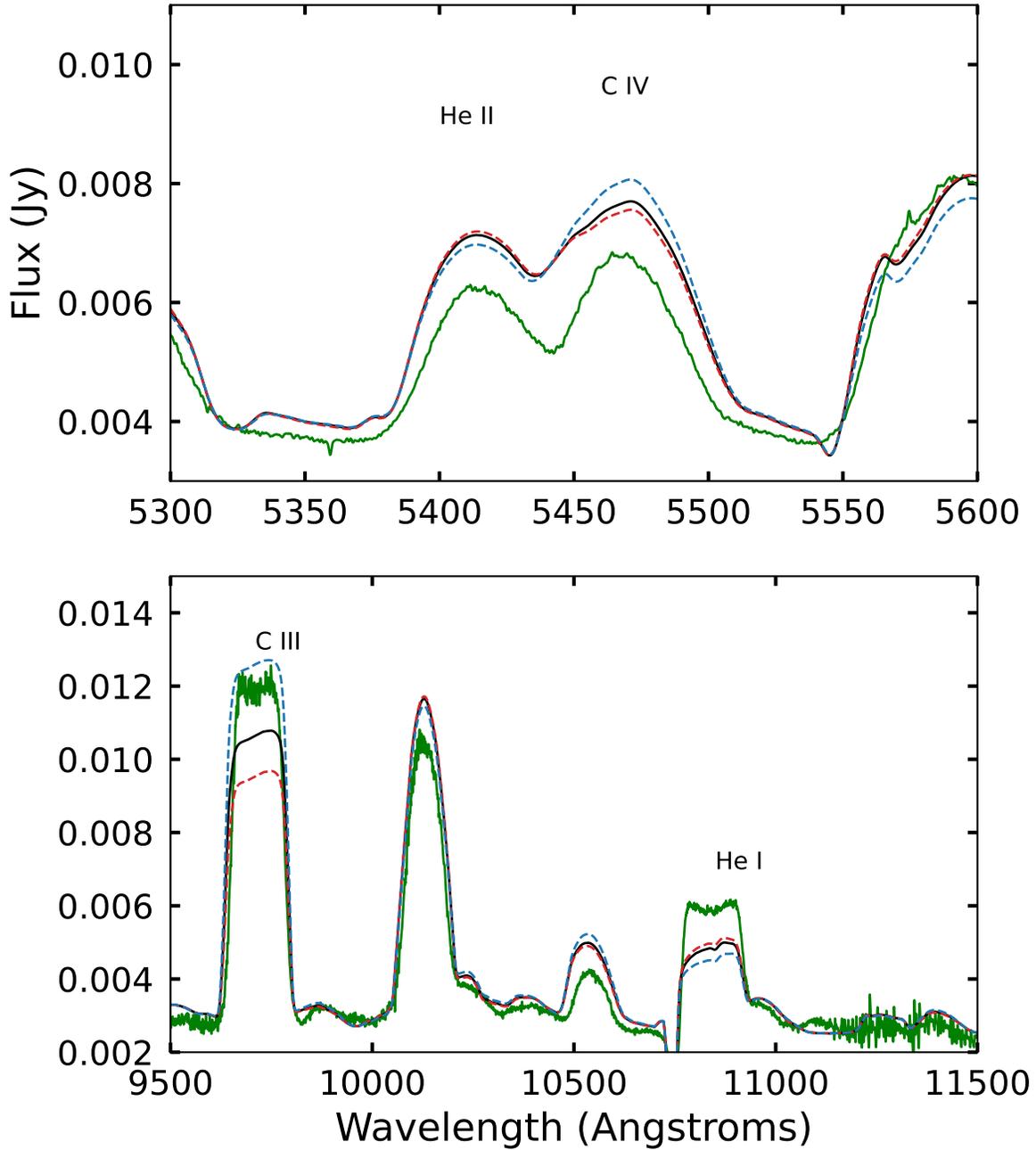}
\caption{C/He uncertainty models.  The carbon/helium line pairs, C\,{\sc iv} $\lambda$5473/He\,{\sc ii} $\lambda$5411 and C\,{\sc iii} $\lambda$9710/He\,{\sc i} $\lambda$10830, used to determine the uncertainty of the model's C/He abundance.  BAT99-8's spectrum (green) and its' best model ($X(\text{C})/X(\text{He}) = 0.83$, black) with the lower uncertainty limit ($X(\text{C})/X(\text{He}) = 0.75$, red dashed) and upper uncertainty limit ($X(\text{C})/X(\text{He}) = 1.08$, blue dashed) scaled by 0.927 for the top plot and 1.085 for the bottom plot.  The lower C/He uncertainty was stopped before the C\,{\sc iv} $\lambda$5473 and He\,{\sc ii} $\lambda$5411 lines flipped flux strengths or had the same flux strength as that would not be representative of the star.  This resulted in the lower C/He limit being very similar to the best model abundances.  However, both the upper and lower C/He uncertainty at the C\,{\sc iii} $\lambda$9710 and He\,{\sc i} $\lambda$10830 lines do not represent the correct ratio, whereas the best model is equally too low for both the C\,{\sc iii} and He\,{\sc i} lines giving a ratio that is representative of the star.
\label{Fig-carbon-error}}
\end{figure}

\begin{figure}[ht!]
\plotone{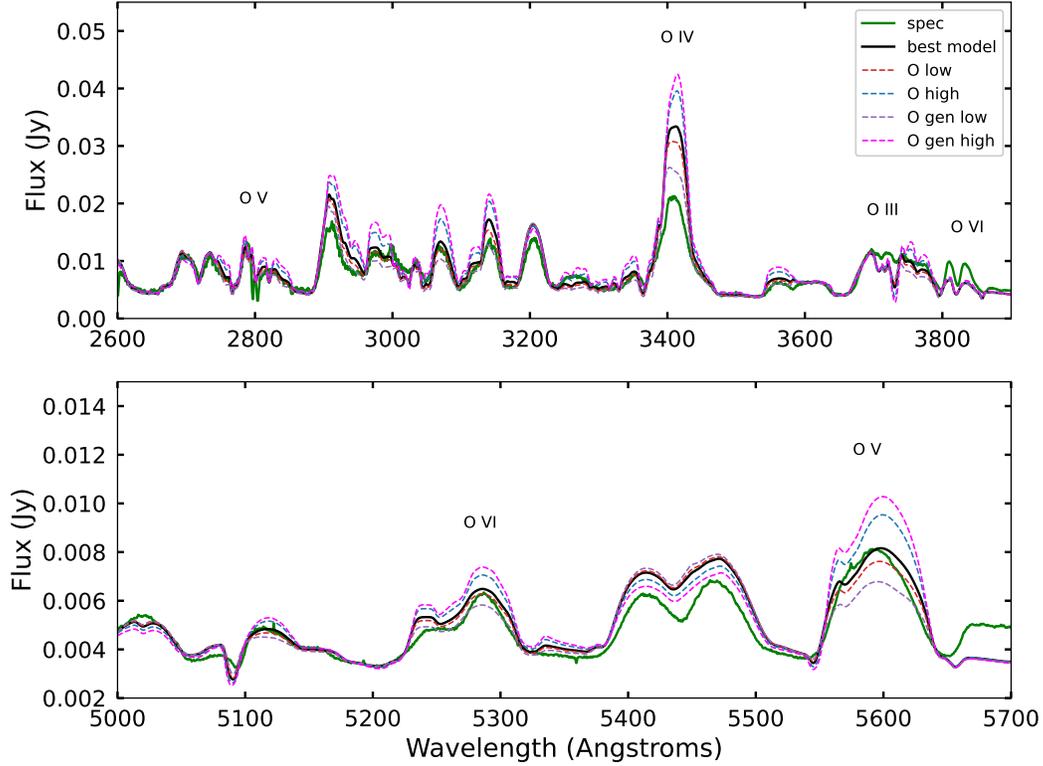}
\caption{Oxygen uncertainty models.  BAT99-8 (green) with its best model ($X(\text{O}) = 0.077$, black) and the ``generous'' and realistic oxygen abundance uncertainty models scaled by 0.927 in the top plot and by 0.930 in the bottom to match the spectrum's ``continuum.''  The realistic uncertainty models (lower: $X(\text{O}) = 0.06$, red; upper: $X(\text{O}) = 0.14$, blue) were adopted as the uncertainties on the oxygen abundance.  The ``generous'' uncertainty models (lower: $X(\text{O}) = 0.04$, purple; upper: $X(\text{O}) = 0.21$, magenta) were used to show the limits of the oxygen abundance in the star.  The realistic uncertainties were determined primarily using the O\,{\sc vi} $\lambda$5291 (recombination line) and O\,{\sc v} $\lambda$5596 lines.  The O\,{\sc iv} $\lambda$3400 line was used to determine the lower ``generous'' oxygen abundance as when we get close to fitting the $\lambda$3400 line, the other model oxygen lines become too low to fit the spectrum.  The upper ``generous'' limit was found by making all of the model oxygen lines too high for the spectrum.  The O\,{\sc iv} $\lambda\lambda$3811,3834 doublet is not reproduced well in the spectrum (see Appendix D).
\label{Fig-oxygen-error}}
\end{figure}

\begin{figure}[ht!]
\plottwo{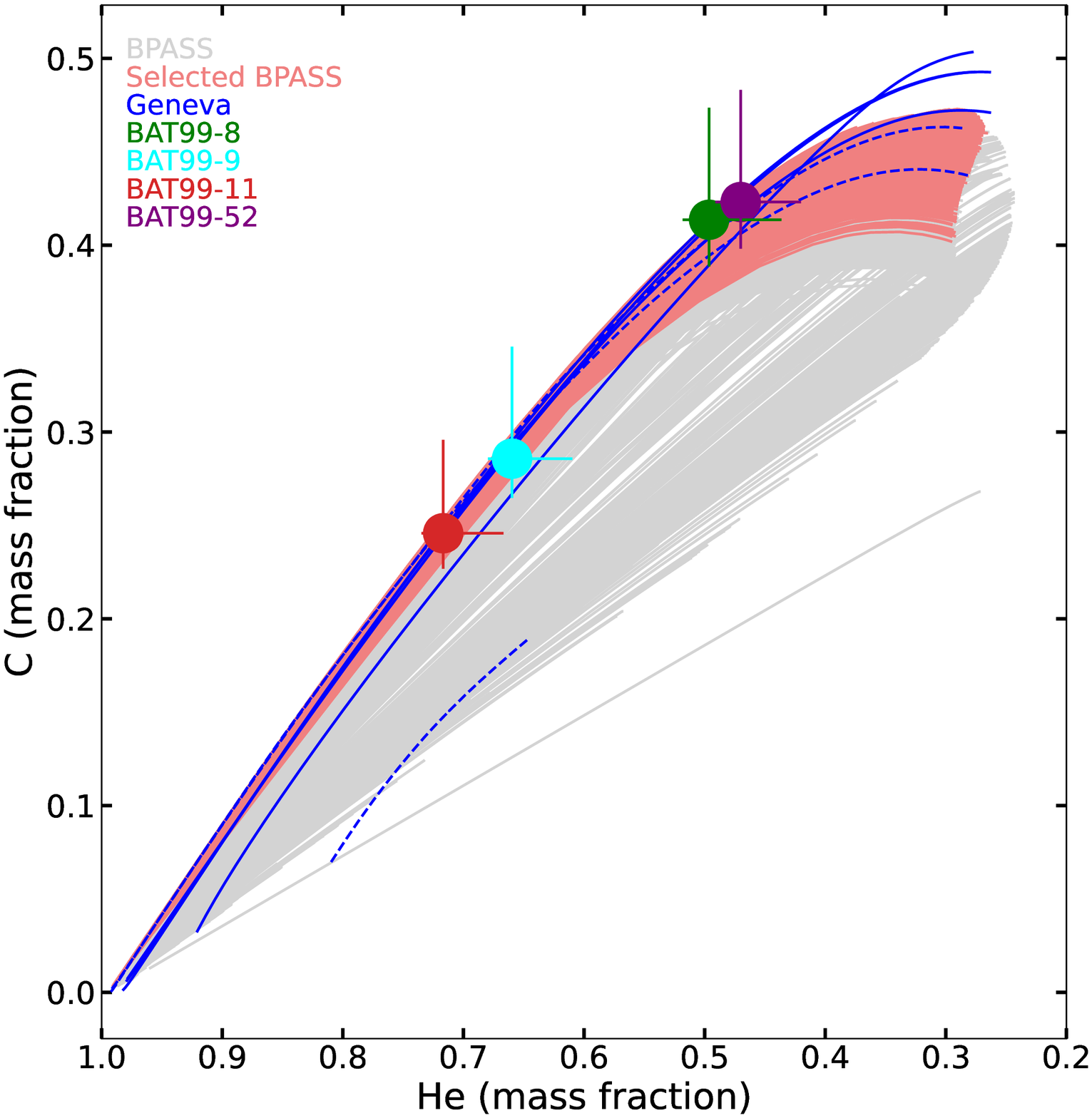}{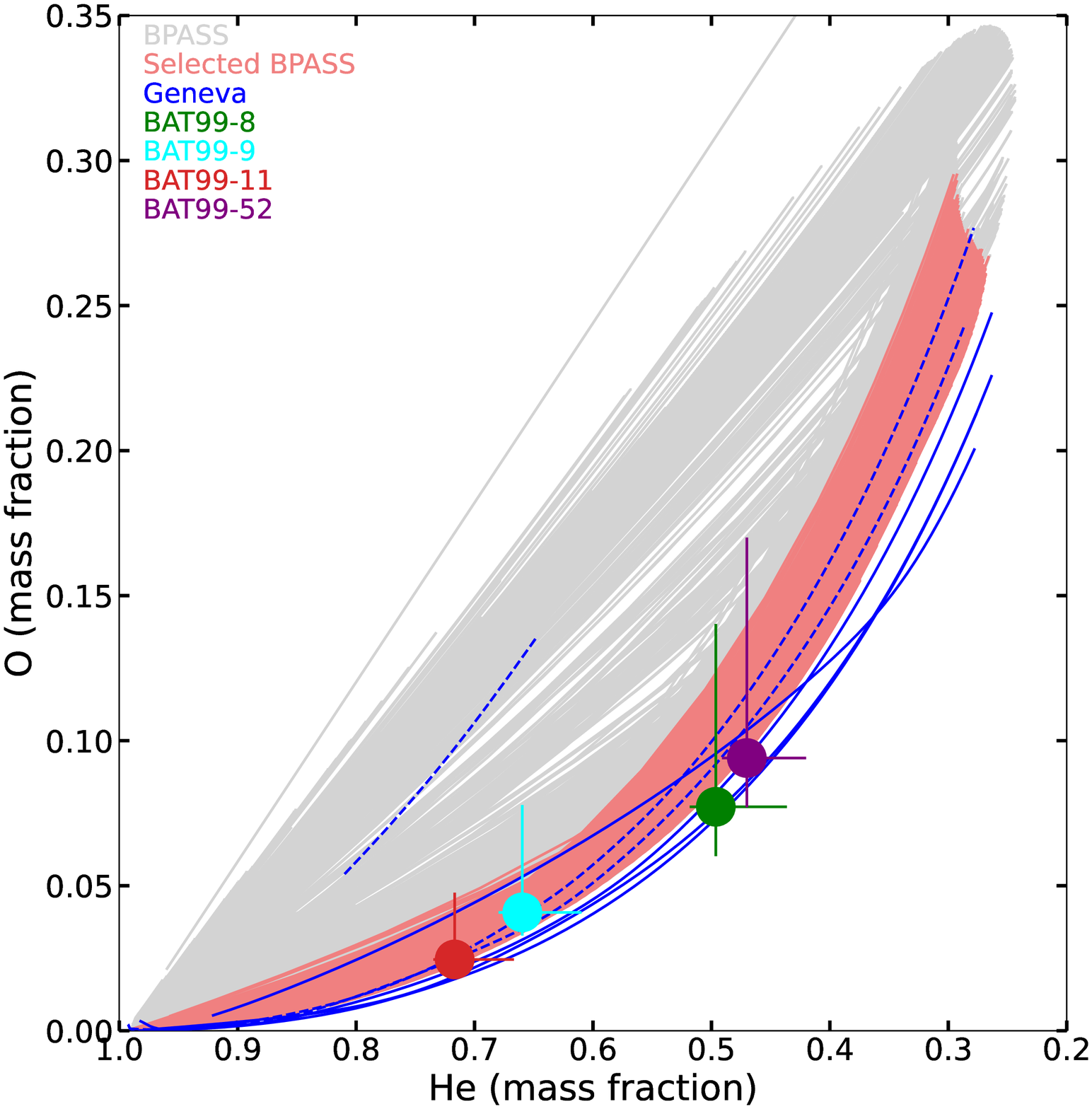}
\caption{Helium vs. Carbon and Oxygen mass fractions.  The BPASS models (gray; coral for models with similar chemical abundance ratios as the WC-type stars), Geneva solar models (solid blue), and Geneva LMC models (dashed blue) show the evolutionary tracks of the models containing a WC phase.  The 4 WC stars are represented as the dots (BAT99-8: green, BAT99-9: cyan, BAT99-11: red, and BAT99-52: purple).  The mass fractions for He, C, and O of the four WC stars are in agreement with both the Geneva single-star and BPASS binary evolutionary models.  In fact, the stars follow the evolutionary tracks, e.g. BAT99-9 and BAT99-11 appear to be less chemically evolved than BAT99-8 and BAT99-52 (have a higher C and O abundance).
\label{Fig-He-C-O}}
\end{figure}

\begin{figure}[ht!]
\plotone{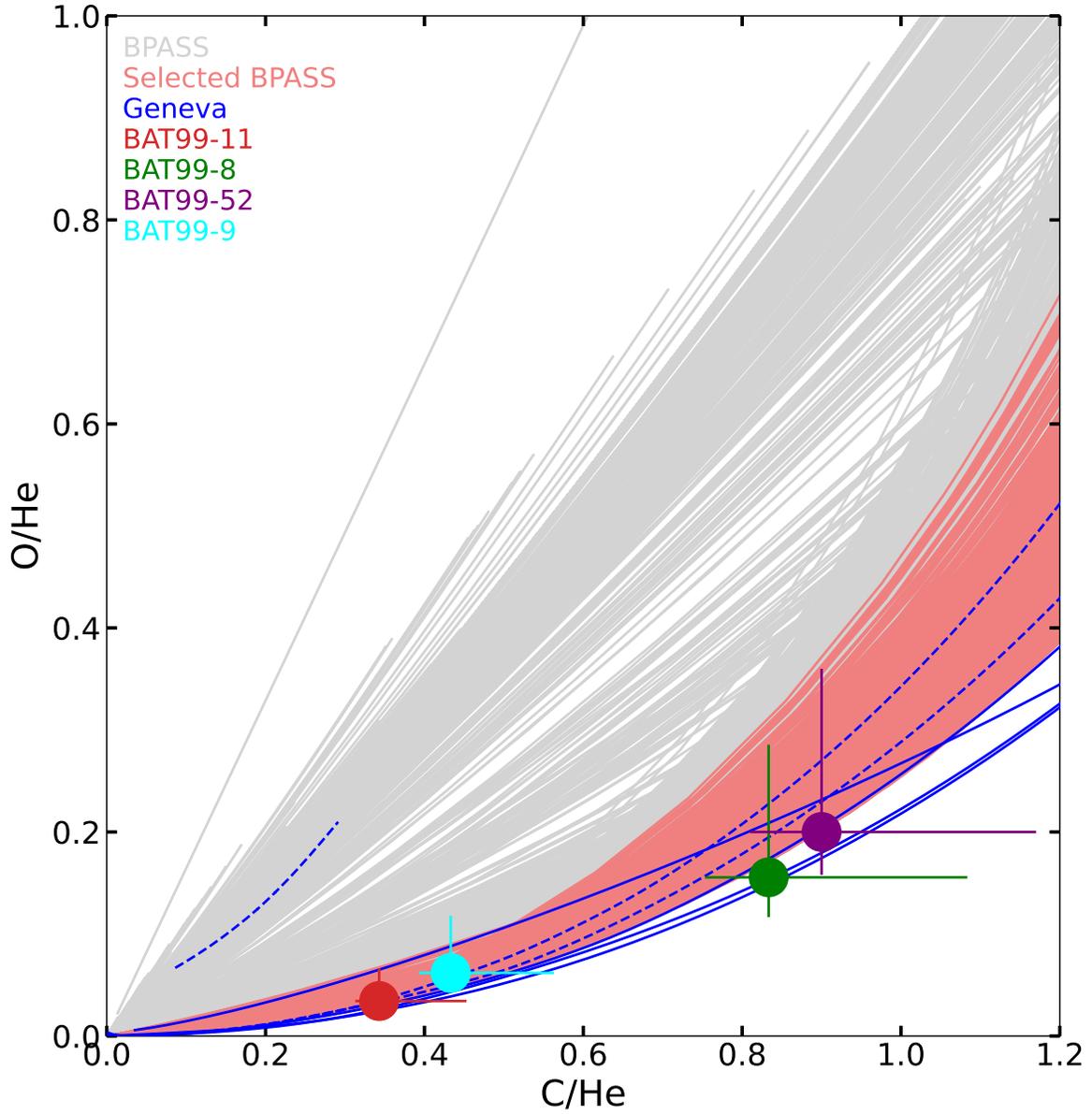}
\caption{O/He vs. C/He plot.  BAT99-8 (green), BAT99-9 (cyan), BAT99-11 (red), and BAT99-52 (purple) are shown as dots on the plot. The BPASS (gray), Geneva LMC (dashed blue), and Geneva solar metallicity (solid blue) models show the evolutionary tracks for the WC phase.  The 4 WC stars lie directly on the Geneva solar metallicity models, specifically the 60, 85, and 120 $M_\odot$ progenitor solar metallicity models. The 40 $M_\odot$ progenitor solar and 85 and 120 $M_\odot$ progenitor LMC metallicity models are in the region but a have higher O/He values.  The Geneva model with a much higher O/He than the stars for a given C/He value is the 60 $M_\odot$ progenitor LMC metallicity model.  The BPASS models span a larger range of O/He values, so only about a third of their models (coral) are near the WC-type stars.
\label{Fig-chemcial_abundance}}
\end{figure}

\begin{figure}[ht!]
\plottwo{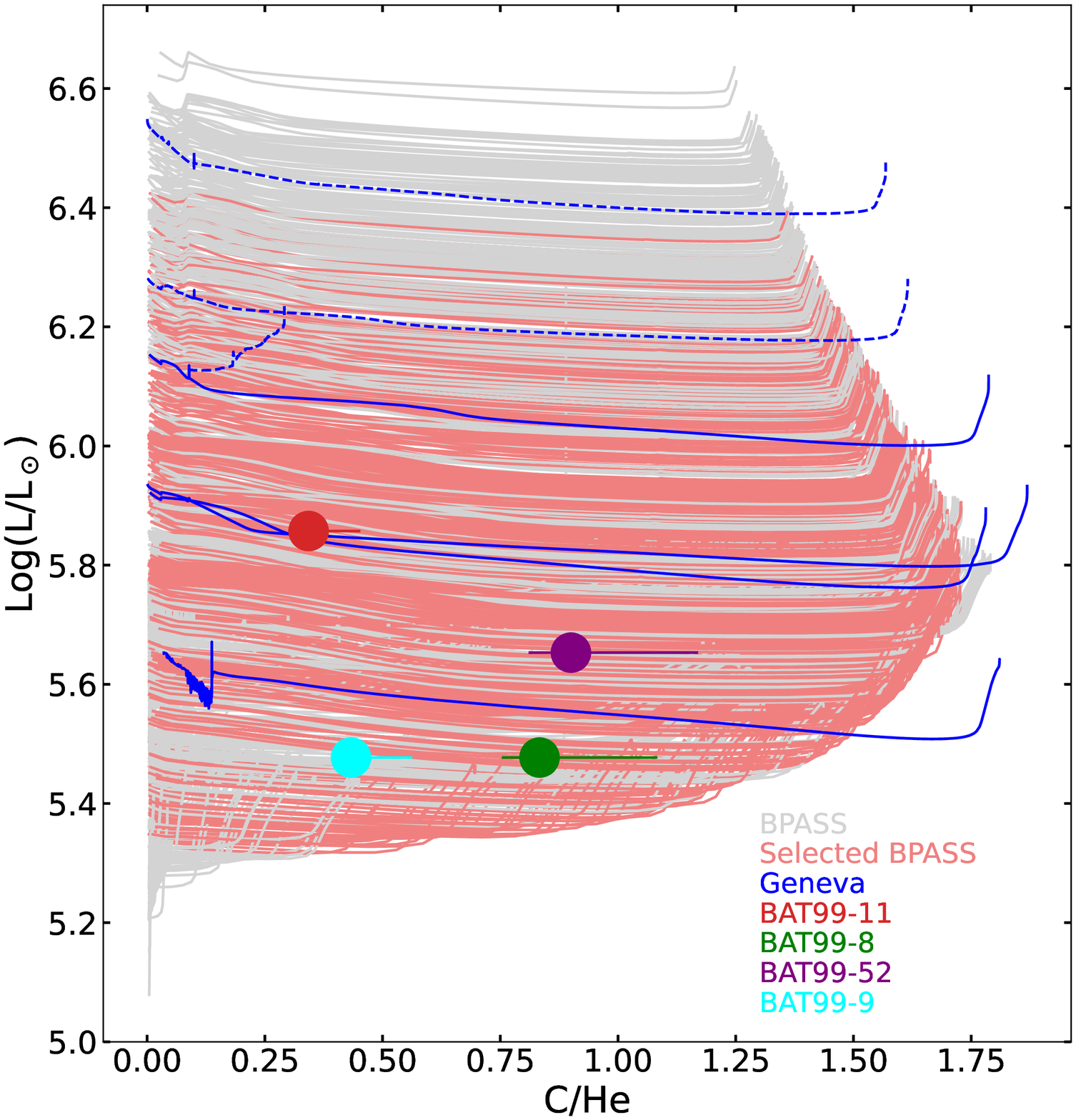}{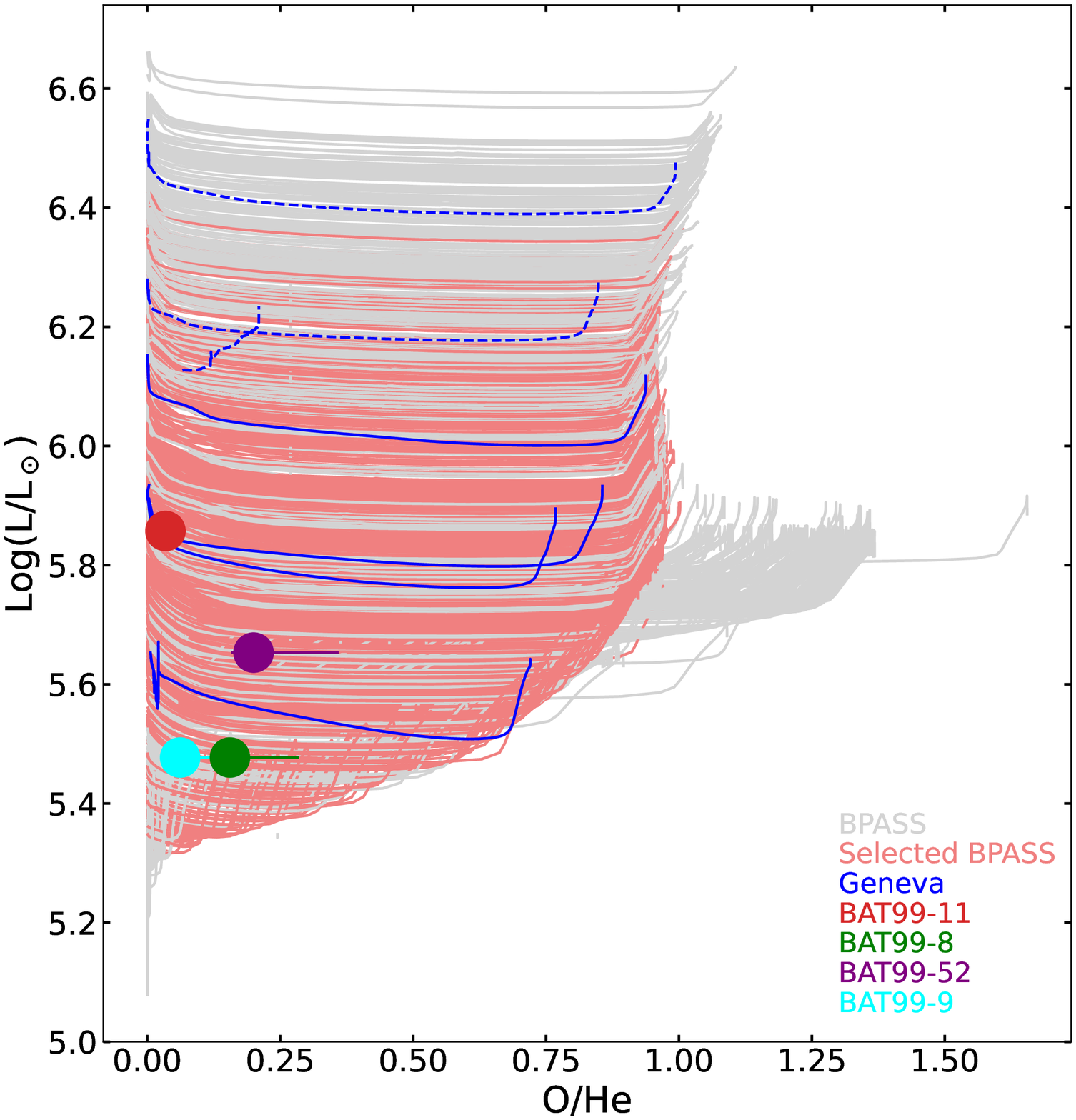}
\caption{Luminosity vs. chemical abundance mass ratios C/He and O/He.  The BPASS models (gray), Geneva solar models (solid blue), and Geneva LMC models (dashed blue) show the evolutionary tracks of the models containing a WC phase.  The 4 WC stars are represented as the dots (BAT99-8: green, BAT99-9: cyan, BAT99-11: red, and BAT99-52: purple).  While the Geneva LMC metallicity single-star models do not intercept with the WC stars, the BPASS binary and Geneva solar metallicity models do.  The progenitor mass 40$M_\odot$ Geneva single-star model is the model closest to BAT99-8 and BAT99-9.  The BPASS models that were near the WC-type stars in Figure~\ref{Fig-chemcial_abundance} are displayed in coral.
\label{Fig-L-CHe-OHe}}
\end{figure}

\begin{figure}[ht!]
\plotone{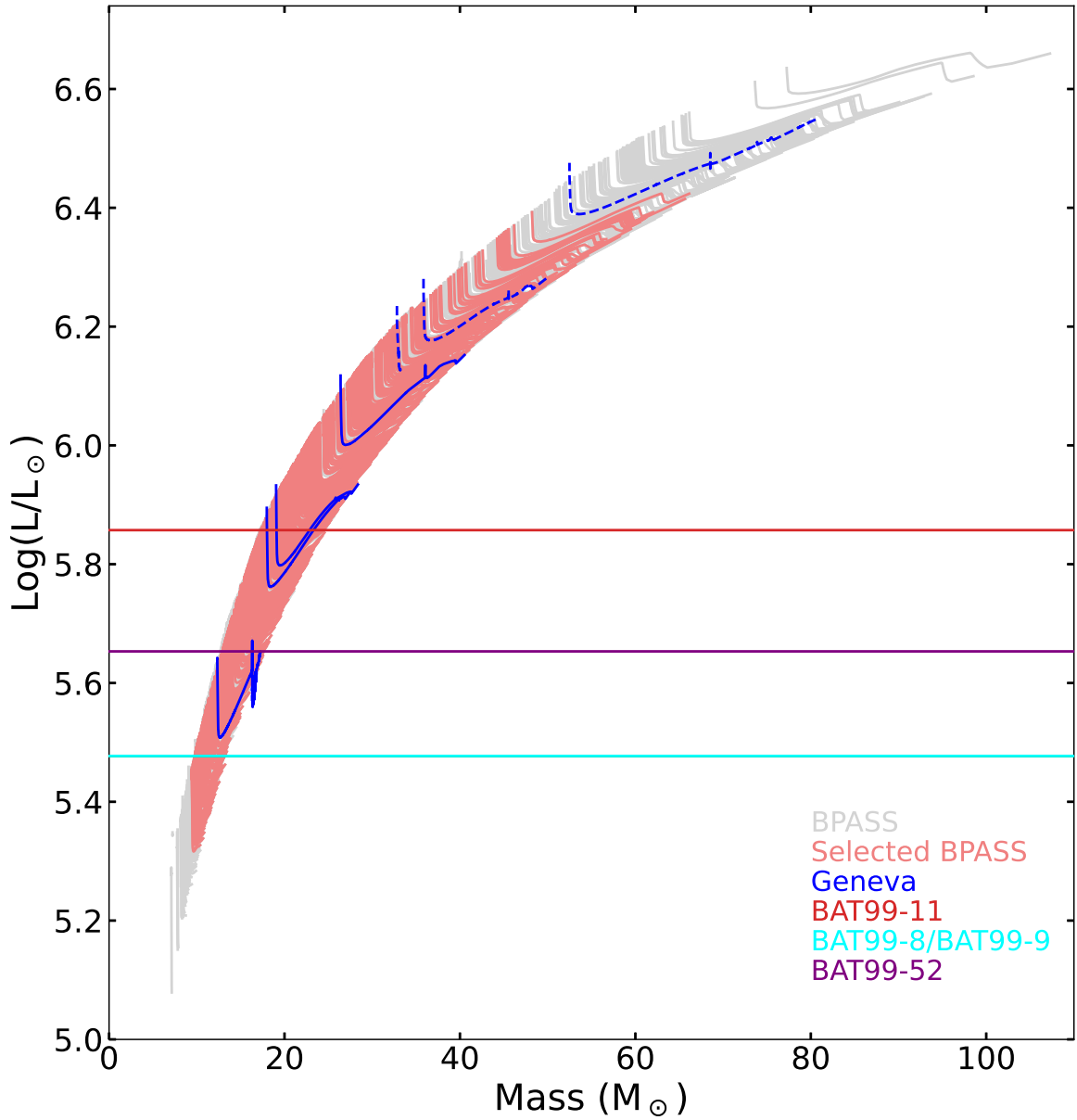}
\caption{Luminosity vs.\ WC mass.  The 4 WC stars are indicated by lines across the plot (BAT99-8/BAT99-9: cyan, BAT99-11: red, and BAT99-52: purple).  BAT99-8 and BAT99-9 have the same luminosity so are represented by one line.  The BPASS models (gray; models near WC-type stars in Figure~\ref{Fig-chemcial_abundance} are coral)  and Geneva models (Geneva metallicity: blue solid; LMC metallicity: blue dashed) have the same luminosity-mass relation, indicating that the type of evolution might not matter for WC-type stars.
\label{Fig-logL-m}}
\end{figure}

\begin{figure}[ht!]
\plottwo{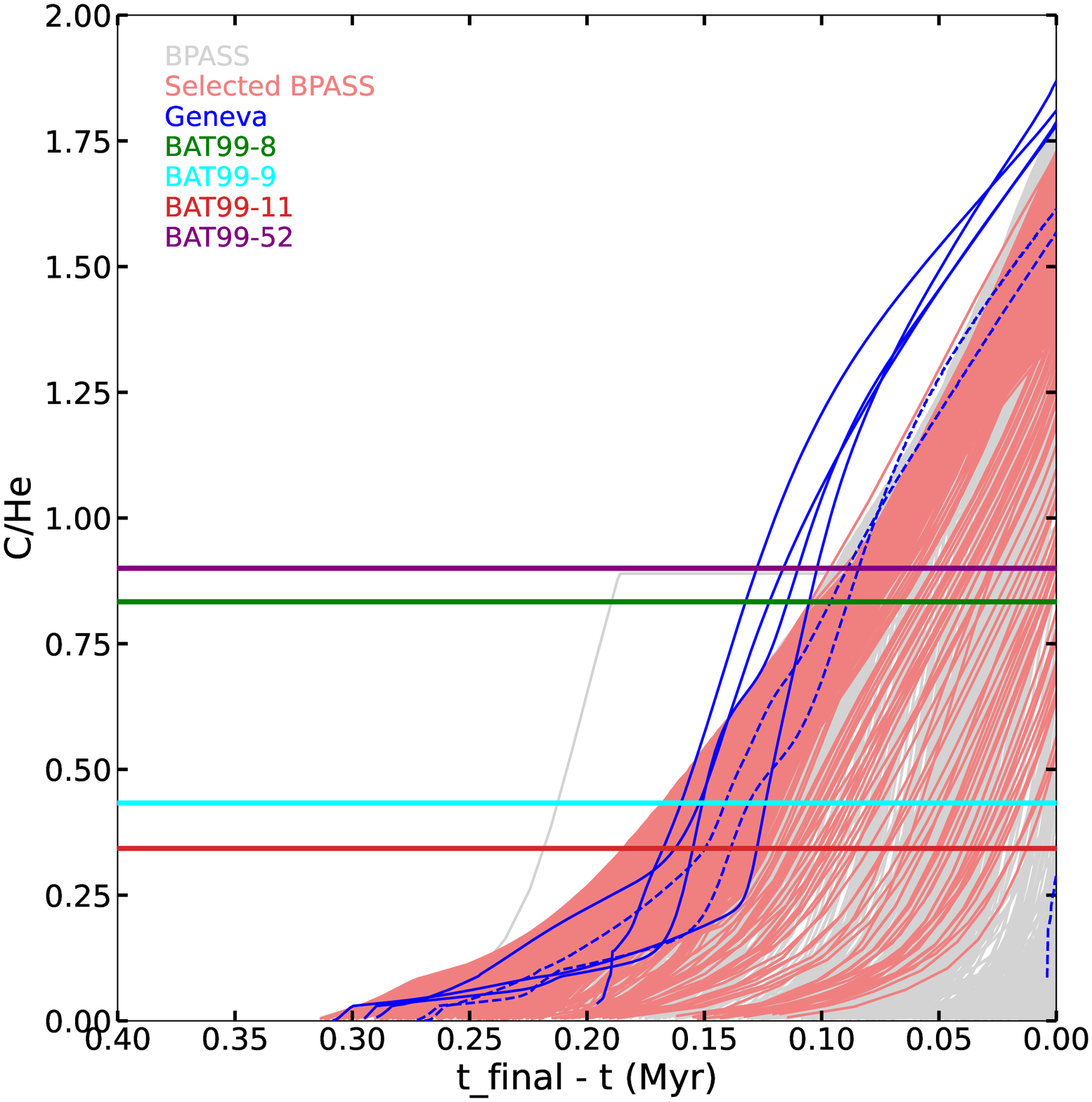}{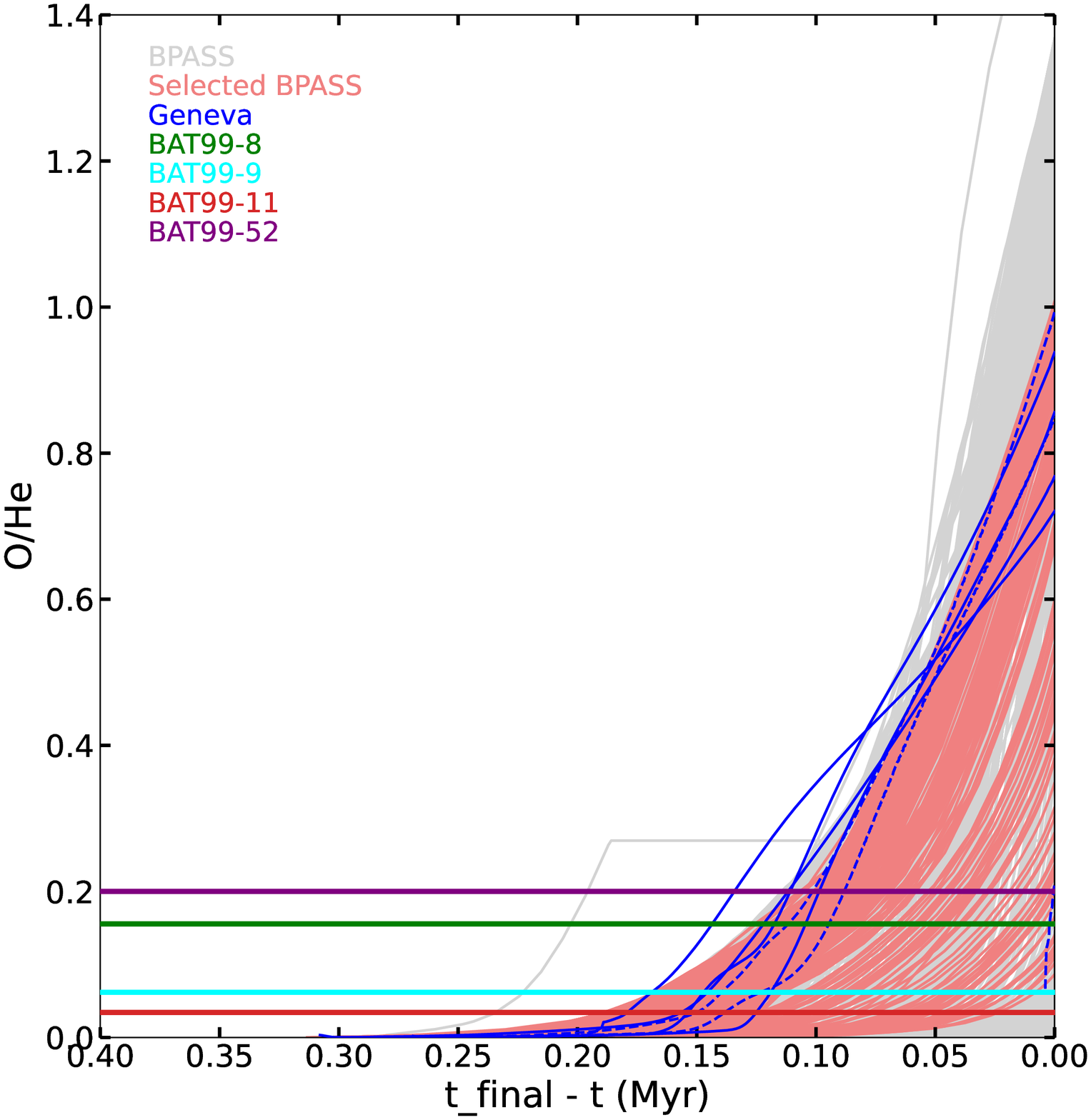}
\caption{Chemical abundance mass ratios, C/He and O/He, vs. time before the star goes supernova. BPASS (gray), BPASS selected for being near the WC-type stars in Fig~\ref{Fig-chemcial_abundance}(coral), Geneva solar (solid blue), and Geneva LMC (dashed blue) models of the WC phase are compared to the WC stars which are represented as horizontal lines since their age is unknown (BAT99-8: green, BAT99-9: cyan, BAT99-11: red, and BAT99-52: purple).  The times at which the stars' C/He and O/He overlap the evolutionary models occurs within a few hundredths of a Myr from each other. 
\label{Fig-t-CHe-OHe}}
\end{figure}

\begin{figure}[ht!]
\plottwo{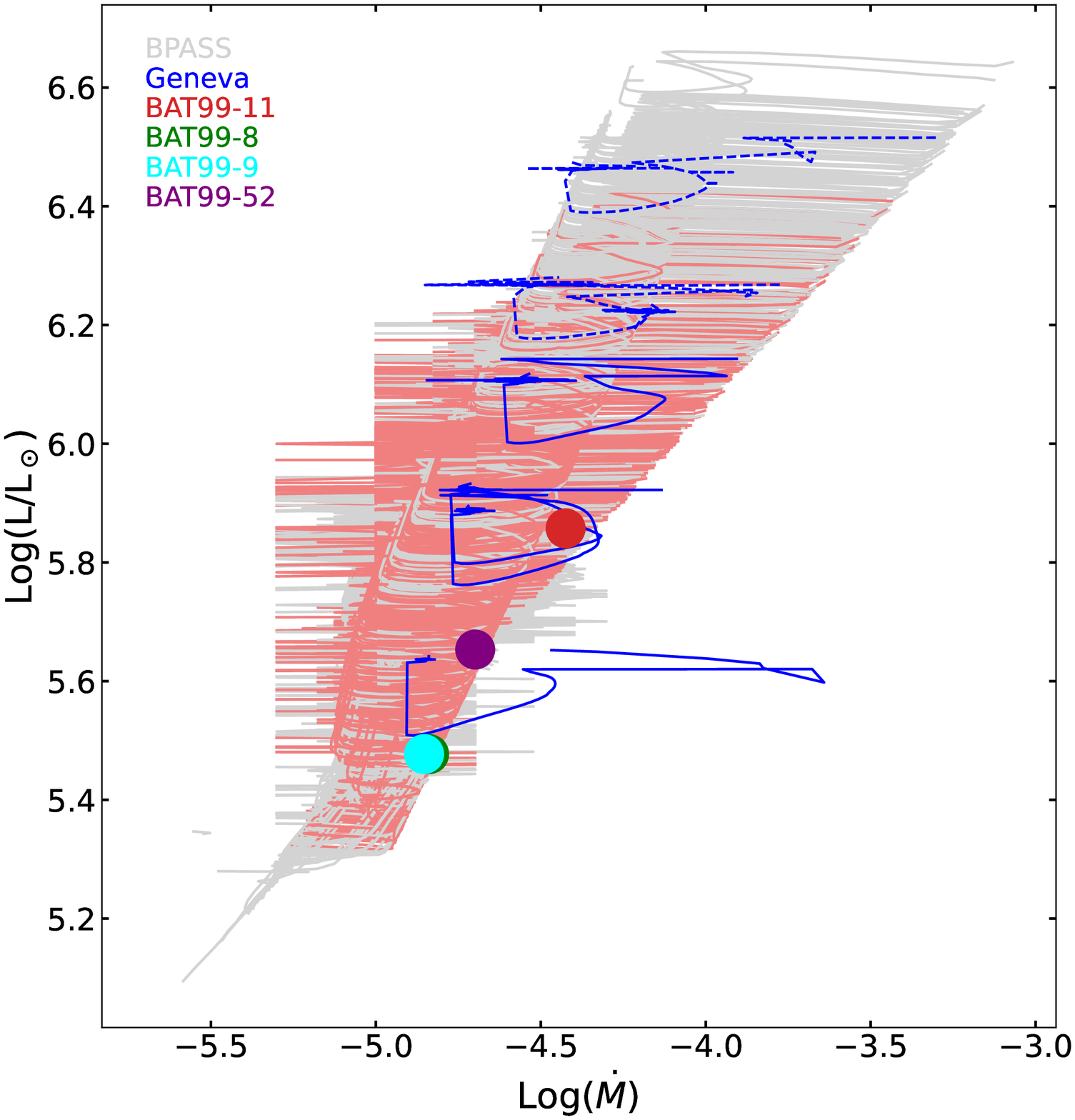}{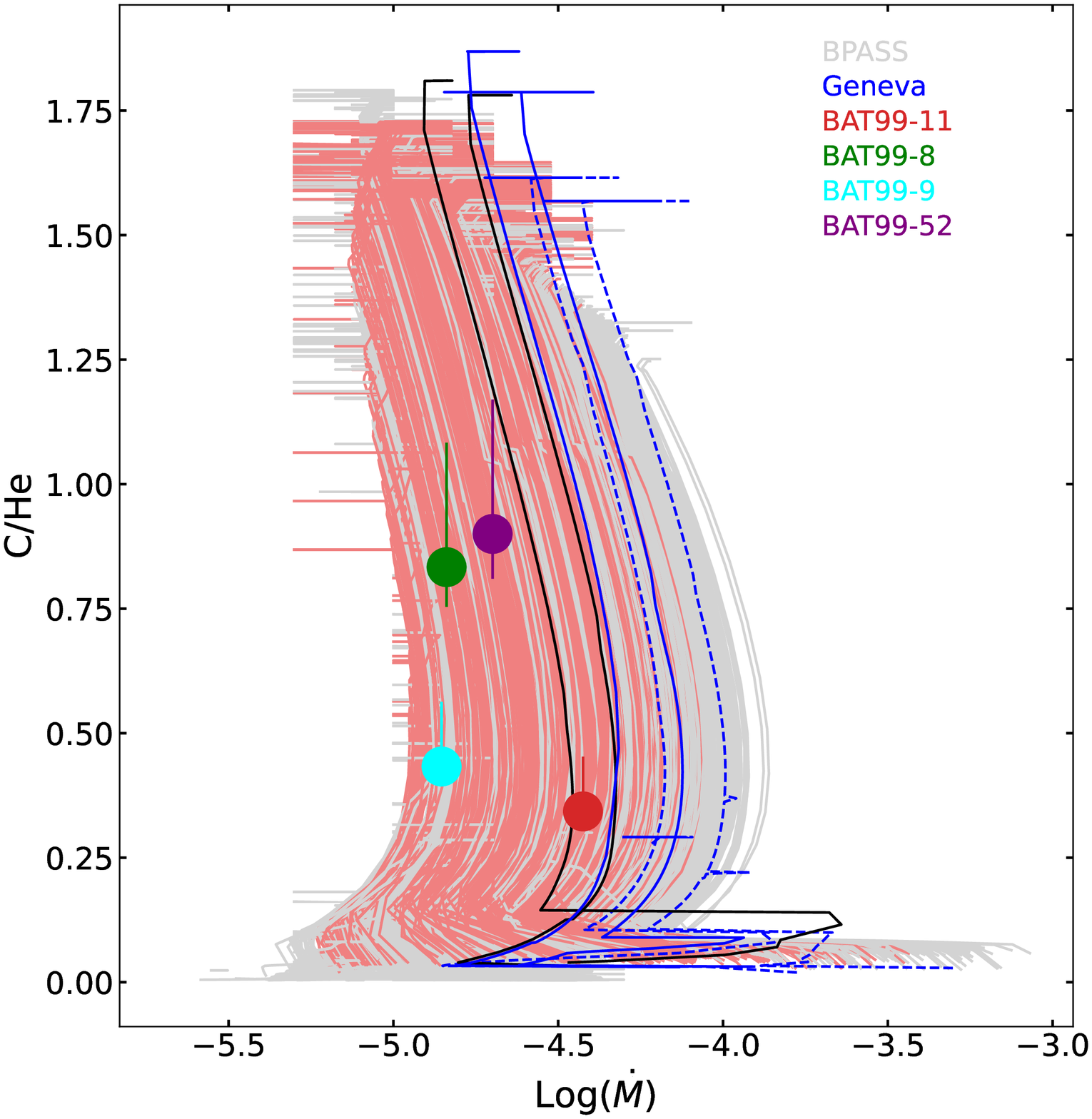}
\caption{Luminosity and Chemical abundance ratio, C/He, vs. Mass-Loss Rate.  BPASS (gray), BPASS selected for being near the WC-type stars in Fig~\ref{Fig-chemcial_abundance}(coral), Geneva solar (solid blue), and Geneva LMC (dashed blue) models of the WC phase are compared to the four WC stars represented as dots (BAT99-8: green, BAT99-9: cyan, BAT99-11: red, and BAT99-52: purple).  The stars agree with the Geneva solar metallicity models and the BPASS models. 
\label{Fig-massloss}}
\end{figure}

\clearpage
 
\appendix

Systematic errors are a major concern in modeling stellar atmospheres. They occur because
of inaccuracies in the modeling, because of assumptions used in the modeling, and because of biases of those doing the analyses.  Below we address four concerns:
\begin{enumerate}
\item
The insensitivity of the model to the adopted stellar radius.
\item
Hydrodynamics and radiative driving.
\item
Clumping.
\item
The inability to fit \osixdoub.
\end{enumerate}

\blankline
\section{The insensitivity of the model to the adopted stellar radius}

The winds of WC stars are optically thick, and as a consequence the radius of the hydrostatic core is not well constrained by spectral modeling \cite[e.g.,][]{1991IAUS..143...59H,Najarro1997,Hamann2004} -- its determination requires detailed hydrodynamic modeling that is beyond the scope of the current paper. 

The difficulties in calculating the hydrodynamic structure are manyfold. In particular,
calculations show that the iron opacity bump is crucial for driving the winds in early-type WN stars and WC stars \cite[e.g.,][]{NL02_mdot,2018A&A...614A..86G,Grafener2005,2020MNRAS.491.4406S}. This opacity can push the star above the effective Eddington limit\footnote{Defined using all opacity sources, not just that due to electron scattering which is used to define the classical Eddington limit.} which can trigger weak convection, density inversions, a wind to be launched, or cause the radius of the star to inflate  \citep{1999PASJ...51..417I,2012A&A...538A..40G}. A further complication is that there are instabilities in a medium when it is close to the Eddington limit \citep[e.g.,][]{S01_edd_instab}. Handling such structures will require time-dependent  3D radiation hydrodynamics.  Since we cannot derive a stellar radius, we cannot determine effective temperatures for comparison with stellar evolution calculations (which also suffer from the aforementioned difficulties).

Surprisingly, the determination of the stellar luminosity appears to be much more robust -- the luminosity is one of the principal factors controlling the ionization structure of the wind, and hence the relative strength of lines belonging to a single chemical element but in different ionization states (e.g., C\,{\sc ii}, C\,{\sc iii}, C\,{\sc iv}). We do stress, however, that the flux we can observe (ie., that longward of the Lyman edge at 912\,\AA) constitutes only $\sim 30\%$ of the emitted energy for the stars studied in this paper.
%C$^{+}$, C$^{2+}$, C$^{3+}$
\section{Wind hydrodynamics and radiative driving.}

Ideally we would like to compute the structure of the wind from first principles. Such calculations are
presently intractable, but can be done in more limited forms. For example, the clumping structure (see below) can be specified \citep{Grafener2005,2020MNRAS.491.4406S}. It  is also very difficult to get the calculations to give the same qualitative agreement with observations as in models where the velocity law and mass-loss rate are treated as free-parameters, although the self-consistent
model of  \cite{Grafener2005} does come close. As noted by \cite{Grafener2005}, the hydrodynamic solutions are also very time consuming.

As for O stars, \cmfgen\ allows a consistency check between radiative force obtained in the modeling
and that required to drive the wind. Since we cannot determine the stellar gravity, the gravity
and mass are free parameters. However, stellar evolution predicts a relatively tight relation between
mass and luminosity for He-burning WR stars \citep{1989A&A...210...93L}. Unfortunately there is scatter. In the BPASS models the range is of order $\pm$10\% for masses less than 20\,\Msun, and somewhat larger for higher masses. While such a range is small, it could have a crucial influence when the star is very close to the Eddington limit.

In most of our modeling we adopted a smaller set of species and ionization stages than warranted by the density and temperature structure of the atmosphere. With this data set there is a large inconsistency between the radiative force obtained in the modeling and that required to drive the wind.  Considerable improvement was obtained by choosing more species, and more ionization stages, and by
making other improvements in some of the atomic data (e.g., by including intercombination transitions)
although in general there are still regions where the radiative force lies below that required to
drive the flow.  A summary of the model ions, the
number of super levels, and number of full levels
for the small data set is provided in Table~\ref{tab_atom_small}.
In Table~\ref{tab_atom_im} we provide a list of changes in the atomic models between the ``small" and ``intermediate" atomic models. Additional species that were included in the
``large" model are listed in Table~\ref{tab_atom_large}.
\ion{Ni}{9} was also included in this table since additional
levels were used.

\begin{figure}  
\includegraphics[width=1.0\linewidth, angle=0]{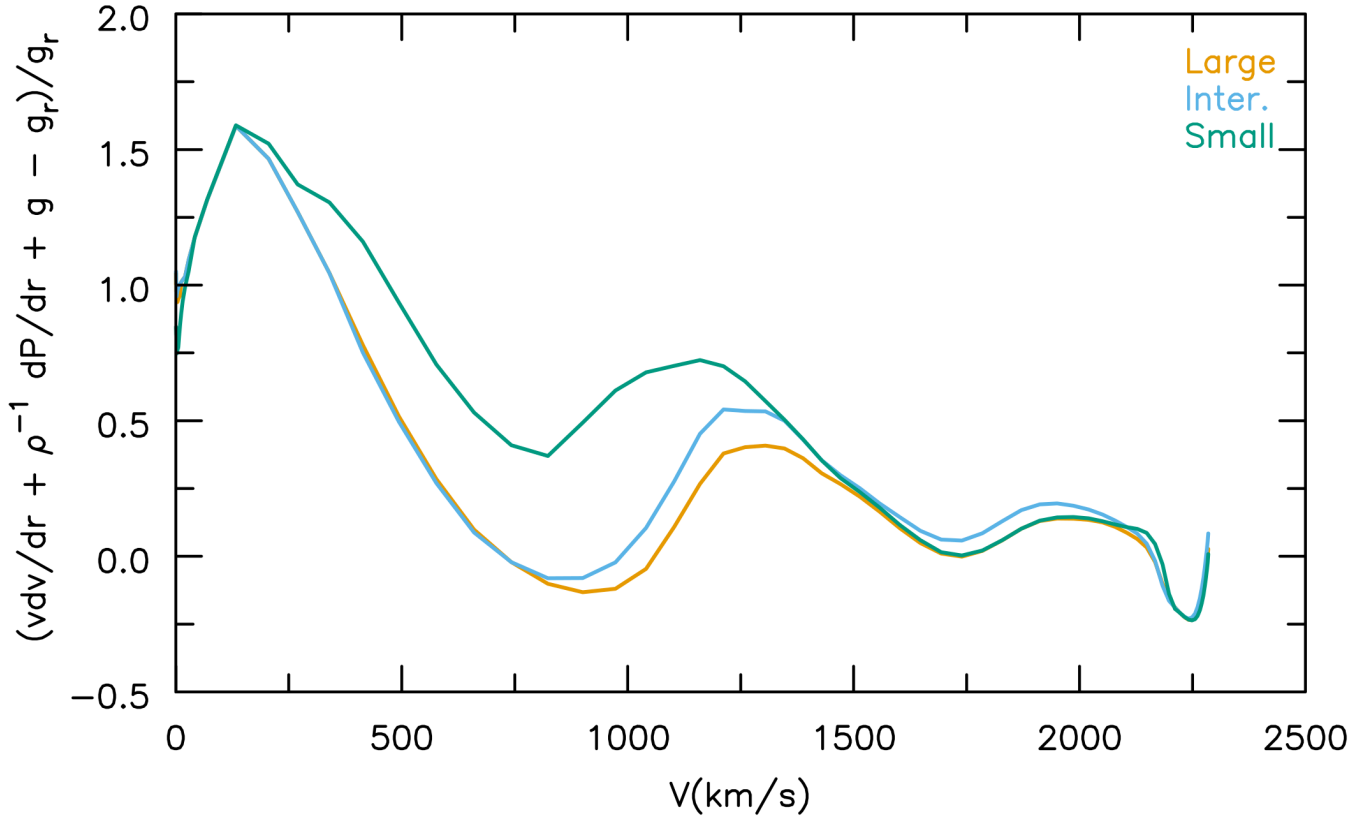}
\caption{Check on the consistency of the radiative force needed to drive the
mass-loss for the assumed clumping and velocity law. The most significant discrepancy occurs below 500\,\kms. The small case refers to the atomic data used to compute the models used in this study (4297 levels, 123 ionization states, 373\,238 lines); the
intermediate case uses ionization stages and levels (5290 levels, 130 ionization states, 844, 731 lines), while the largest set had 6209 levels, 20 chemical species, 149 ionization states, and treated 1\,210\,981 lines. All models had 18 or 19 chemical species: H, He, C, O, N, Ne, Na, Mg, Al, Si, P, S, Cl, Ar, Ca, Cr, Mn, Fe, and Ni.}
\label{fig_hydro}
\end{figure}

While the revised atomic model has a strong influence on the radiative force
(Fig.~\ref{fig_hydro}), it  has only a minor influence on the predicted spectrum. In the optical  ($\lambda$ 3000 to 10000\,\AA) the average flux (computed in pixel space without
regard to resolution) differed by less than 3\%, and the statistical variation between the
spectra was less than 4\% (1 sigma). The most significant changes occurred for the
\ovtrip\ / \oiiising\ blend,  and \ciiinirtrip\ which were enhanced by roughly 20\% in model computed with the revised atomic model. Similarly in the UV changes, except in some deep P~Cygni absorptions, were less than 10\%, and the statistical variation was less than 5\%.
The models used  have a luminosity of $3\times 10^{5}\,\Lsun$
and a stellar mass of 15.0\,\Msun\ (the mass derived from the formula of \citealt{1989A&A...210...93L} is 14.8\,\Msun).

In order to drive a stellar wind it is necessary that the radiative force increases as we pass through the sonic point\footnote{At the sonic point the wind speed matches the isothermal sound speed. In thermal winds it is also a critical point of the flow -- physically this
occurs because material above the sonic point cannot communicate to material below
the sonic point by sound waves. In radiative flows, in which the driving force
depends on the velocity gradient, there is still debate about whether the critical
point occurs at the sonic point, or at higher velocities \citep[e.g.,][]{L07_wind_struct}. The
latter is the case when the force has an explicit dependence on $dv/dr$ as in the Sobolev approximation
\citep[e.g.,][]{CAK1975,PPK86_winds}.}.
Because of the Fe-opacity bump, this can place limits on the temperature and density structure at the sonic point \cite[e.g.,][]{NL02_mdot,Grafener2005}. Indeed, by adjusting the radius of our models we can improve the consistency in the hydrodynamics around the sonic point. Again, this does not lead to significant spectral changes, and hence does not lead to significant changes in the stellar luminosity, or abundances.

Because the inner wind is optically thick we have no direct constraints on the velocity law
in a large portion of the wind. However, in principle, we can constrain $V(r)$ in the outer wind through  profile fitting. Unfortunately even this is difficult because of severe blending, and because the profiles
also depend on the variation of the clumping factor with radius.

\cite{2020MNRAS.491.4406S} derived self-consistent mass loss rates for galactic WC stars from hydrodynamic models assuming a volume filling factor of 0.02 (set to 1 in the inner region). For a $\log (L/\Lsun)/(M/\Msun)$ ratio of 4.3 (corresponding to $L=3 \times 10^5 L_\odot$, $M=15\,\Msun$) a mass-loss rate of $7 \times 10^{-6}\,\Msunyr$ was derived. This is roughly a factor of 2 lower than our derived mass-loss rates for BAT99-8 and BAT99-9, but the discrepancy would be even larger if we allowed for the lower metallicty of our WC LMC stars.

\section{Clumping}

It is well known that the winds of WR stars (and O stars) are clumped. The strongest evidence for
clumping in WR stars comes from variability studies \citep[e.g.,][]{1988ApJ...334.1038M,1988PhDT........13M,1999ApJ...514..909L}, and the weakness of the red-shifted electron-scattering wing relative to its associated emission line \cite[e.g..,][]{Hillier1991b}. In WR stars the wings are redshifted because the typical expansion velocities ($\sim$2000\,\kms) are much larger than the thermal motions of the electrons ($\sim 400 (T/10^4)^{0.5}$\,K). If we ignore the thermal motions, every scattering of a photon by an electron in a monotonic flow leads  to a decrease in the photon's energy. A major advantage of clumping is that it reduces the mass-loss rate, and hence it makes it easier for radiation pressure to drive the flow.

In \cmfgen\ we assume the clumping at a given radius to be uniform, and typically use two parameters to describe the radial variation of  clumping -- an  onset velocity and the clumping at infinity. This prescription is completely ad hoc, and is not based on physics. We have very little idea of how the clumping varies with radius/velocity, and it is very difficult to derive constraints using observations. 
We also make a simplistic assumption to solve the radiative transfer equation. In particular we assume that the clumps are much smaller than a photon mean-free path, and that the interclump medium is void. Despite these assumptions and approximations, \cmfgen\ (and similar codes)
are capable of producing theoretical spectra in excellent agreement with observations.

The choice of clumping law can have an influence on the quality of fits. As WR winds are optically thick, both the continuum (especially at optical and longer wavelengths) and lines originate in the wind. Since a strong emission line arises at larger radii than its adjacent continuum, its equivalent width will be affected by radial variations in clumping. Further, the biases introduced by our simplistic approach to clumping will influence line and continuum formation differently.

\begin{figure}  
\includegraphics[width=1.0\linewidth, angle=0]{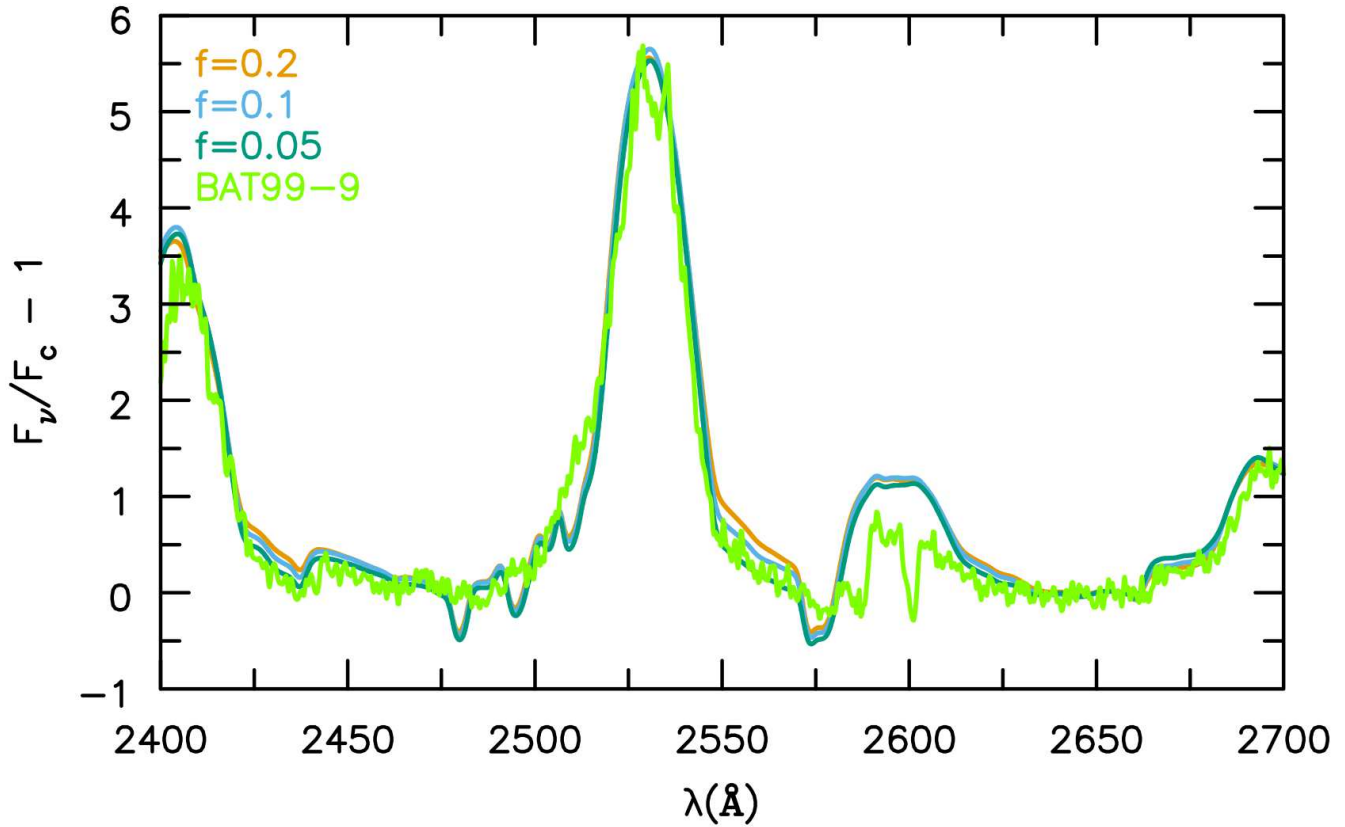}
\caption{Illustration of the influence of different clumping factors on the
\ion{C}{4}\ $\lambda$2530 profile. To remove slight differences in the observed emission line strength
relative to the model we scaled the continuum (seen at 2650\,\AA) to unity, subtracted unity,
and then scaled the observations so that the emission peak was in rough agreement. As
apparent from the plot, a clumping factor of 0.2 leads to an electron scattering wing
(between 2550 and 2575\,\AA) that is too strong compared with the observations.}
\label{fig_es_civ}
\end{figure}

\begin{figure}  
\includegraphics[width=1.0\linewidth, angle=0]{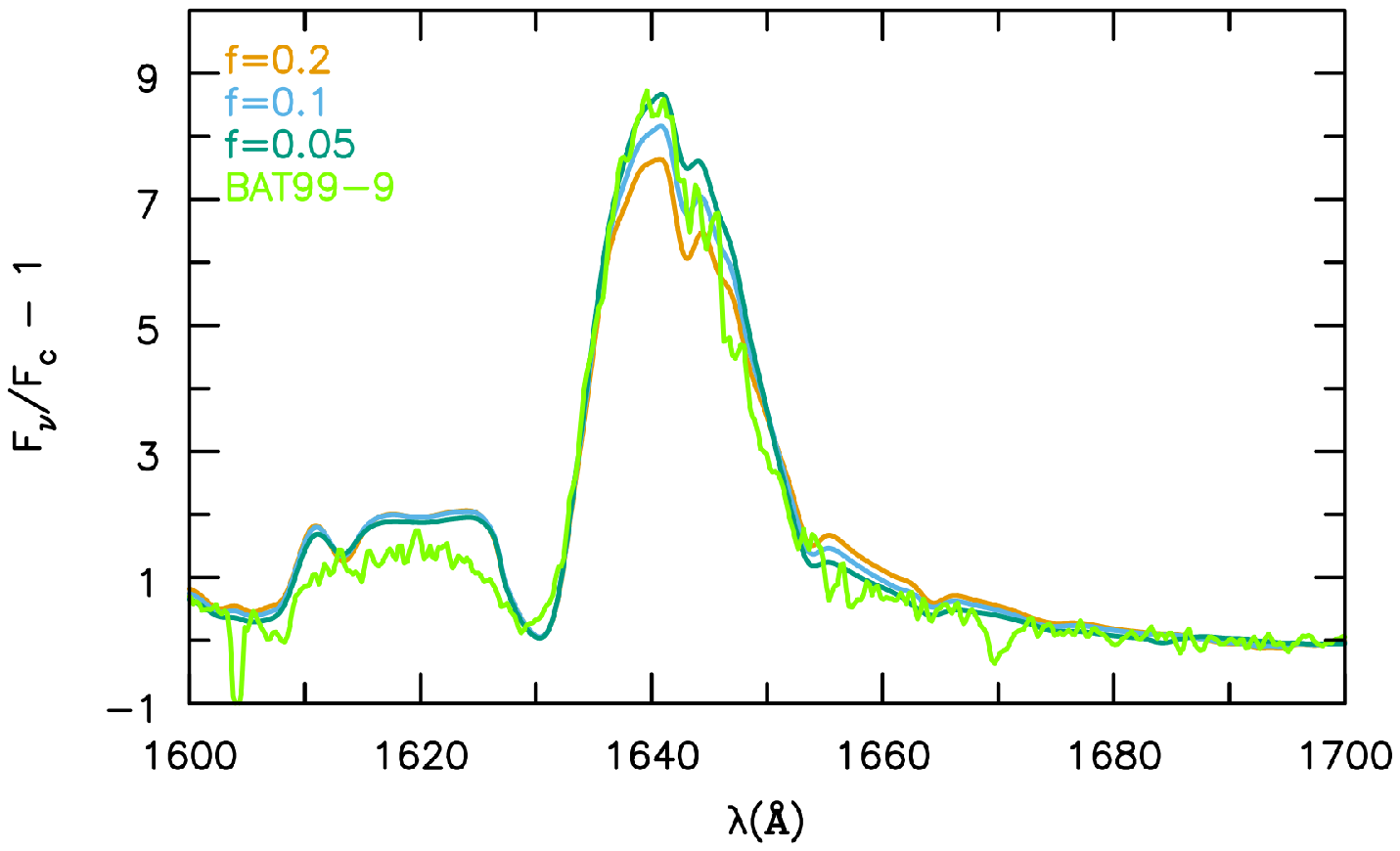}
\caption{Similar to \ref{fig_es_civ} but for \ion{He}{2} $\lambda$1640.
The continua were scaled to match but no additional scaling was performed.  Again
we see that a clumping factor of 0.2 leads to an electron scattering wing
(between 1650 and 1670\,\AA) that is too strong compared with the observations.}
\label{fig_es_he2}
\end{figure}

The influence of clumping is shown in Figs.~\ref{fig_es_civ} and \ref{fig_es_he2} for three values of the volume-filling factor: 0.05 (used in this paper), 0.1, and 0.2. Because of blending and the absence of ``line-free" regions, the influence of electrons is not as easily discerned as in WN stars.  While a value of 0.2 can be ruled out by a comparison between model and observations, it is extremely difficult to place a lower limit on the clumping factor in WC stars. The redshift of the \civres\ transition, which is influenced by its large optical
depth and the need to use a Voigt profile when computing the observed spectrum
\citep{Hillier1989}, also favors a volume-filling factor less than 0.2. 

Because of the complexity of WC spectral formation, identical spectra are not obtained when $\Mdot/\sqrt{f}$ is held constant. However, generally only minor adjustments of one or more parameters (e.g., \Mdot\ or $L$) are needed to obtain better agreement  between models computed with different filling factors.

\section{The inability to fit \oivdoub\ and \osixdoub}

As readily apparent from the modeling, there are a few significant discrepancies between the models and observations. There are many possible reasons for the discrepancies, and tracking down the source of a particular discrepancy is very difficult. With the current modeling two promising sources for the discrepancies are the adopted atomic models, and  our simplistic treatment of radiative transfer in the clumped wind. 

While a close examination of Figs~\ref{Fig-BAT99-8}-\ref{Fig-BAT99-52} reveals many (minor) discrepancies many of these can be regarded as insignificant in the sense that that are likely to be due to issues with the adopted atomic data, or can be removed with a relatively minor change in the adopted stellar parameters. However, there are two major discrepancies between our theoretical spectra and observed spectra which are not easily explained. In particular, the strength of \oivdoub\ is consistently overestimated (factor of 2) while the \osixdoub\ is underestimated (factor of 2 to 4). The latter problem has been seen in earlier modeling of WC stars \citep[e.g.,][]{Hillier1999,Crowther2002,Sander2012}, and is also often seen when modeling WO stars \citep[e.g.,][]{2015A&A...581A.110T}. Coincidently, it is often seen for \nvdoub\ in WNE stars (\ion{O}{6} and \ion{N}{5} are isoelectronic) \citep{2014A&A...565A..27H}. However, it is unclear if the discrepancies have the same cause.

Despite many tests, we were unable to fit  \osixdoub\ while maintaining the fit to other spectral lines. Following the work of  \cite{Grafener2005} we did find that a reduction of clumping in the \osixdoub\  formation region increased its strength (i.e., the equivalent width) significantly. Unfortunately the reduction in clumping reduced the IR fluxes (and to a lesser extent those in the optical), and this enhanced the EW of IR lines so that they were no longer compatible with observation.  

We do not believe a significant increase in luminosity, alone,  would solve the problem. The luminosity we derive is constrained by lines due to multiple ionization stages -- \ion{He}{1}/\ion{He}{2}, \ion{C}{3}/\ion{C}{4}, and \ion{O}{3}/\ion{O}{4}/\ion{O}{5}/\ion{O}{6}. 
We did try enhancing the clumping in the \ion{C}{3}/\ion{He}{1} in the formation zone
(i.e., by decreasing $f$ by roughly a factor of 2) to compensate for the higher luminosity but
none of the models were satisfactory.

It may be possible to get a better fit to  \osixdoub\ by simultaneously increasing the luminosity and altering \Mdot\ and the run of clumping with radius. By increasing the luminosity by 30\%, it is possible to get the strength in close agreement with that observed. In the best model we had $f=0.1$ in the inner wind decreasing to $0.5$ in the outer wind -- the later being necessary to get the lower ionization lines closer to that observed. However in this model \ovtrip\ is now a factor of two stronger than observed, and most of the \ion{C}{4}\ are systematically too strong. It is possible that a more sophisticated treatment of clumping would solve the problem with \osixdoub. In our models there is no shortage of photons -- the line simply has a very large optical depth which inhibits photon escape.

A crucial confirmation of our models is that the \ion{O}{6}\ recombination line ($n=8$ to $n=7$) at $\lambda 5290\AA$ (although blended) matches reasonably well.  Fortunately this line in WO stars is clean, and this will greatly aid in constraining the oxygen abundance in WO stars.  
 
 \cite{Sander2012} were able to fit \osixdoub\ in the galactic WO star WR102 using the
 {\sc P}{o}{\sc WR} code \citep{Grafener2002,Sander2015}. The derived luminosity was $\log L/\Lsun = 5.68$ (for a DM$=12.39$; $\log L/\Lsun=5.58$ when corrected to the Gaia DR2 DM of 12.14, \citealt{2019A&A...621A..92S}), significantly higher than the value of $\log L/\Lsun = 5.45$ (5.0 for DM$=12.14$) found by \cite{2015A&A...581A.110T}. The difference in the results is at least partially attributable to the fitting procedure. While \cite{Sander2012} fitted  \osixdoub\, the fit to lower ionization lines, such as \ovtrip\, was very poor. Conversely, \cite{2015A&A...581A.110T} failed to fit \osixdoub\ but the fit to the lower ionization lines was much better \footnote{For the two galactic stars
analyzed by \cite{Sander2012} the He:C:O mass percentages were 30:40:30 while
 \cite{2015A&A...581A.110T} obtained 14:62:24 for  WR102 and 26:54:21 for WR142.}. In WO stars \osixdoub\ is one of the strongest lines in the optical, while in WC 4 stars it is a minor feature.
 
As noted earlier we also fail to fit \oivdoub, the  2s$^2$~3p~$^2$P$^{\rm o}-2$s$^2$~3d~$^2$D multiplet. The line is blended with the weaker   2s~2p($^3$P$^{\rm o}$)~3s $^4$P$^{\rm o}- $2s~2p($^3$P$^{\rm o}$)~3p $^4$D transition. Two other \ion{O}{4}\ multiplets,  2s$^2$~3s~$^2$S$-$2s$^2$~3p~$^2$P$^{\rm o}$ ($\lambda\lambda 3063-3072$) and 2s~2p($^3$P$^{\rm o}$)~3p $^3$P$-$2s~2p($^3$P$^{\rm o}$)~3d $^2$F$^{\rm o}$ ($\lambda\lambda 3560-3593$) are in better agreement with the observations. An examination of the models and line rates found no obvious cause for the discrepancy.
 
\section{Atomic data}

To do the modeling presented in this paper a great deal of atomic data is needed. 
The data we used comes from a wide variety of sources and is of uneven quality. 
Tests with different atomic data sets generally did not yield significant spectral
changes for lines used in the spectral fitting. This, by itself, does not prove the
accuracy of the atomic data -- different data sets can suffer from the same limitations.
In the present work we include low temperature dielectronic recombination which
typically arises from doubly excited states that lie just above the ionization limit. On the other
hand we neglect high temperature dielectronic recombination (HTDR) that  typically arises from doubly excited states with high $n$ ($\sim100$). For example, in \ion{C}{3}\ HTDR
occurs via  2p $nl$ transitions.  The difficulty of treating HTDR is severalfold -- it occurs
from very large $n$ (and low $l$) values and it can be suppressed at high densities 
\cite[e.g.,][]{1969ApJ...157.1007B, 2018A&A...610A..41K}.
Further, in WR stars the radiation can not be neglected, and this will also suppress HTDR.

Below we list the sources of atomic data used in the WC calculations. All data
are available through the CMFGEN website {\url{www.pitt.edu/~hillier}}.  When changes are
made to the atomic species, a new sub-directory is created.

\subsection{Energy levels}
In general we used energy levels from NIST. For many species we used the atomic data of Bob Kurucz \citep{Kur09_ATD} that is  available through his website \cite{Kur_web}. An advantage of this data is that in most cases the data already contain NIST energies (when available), and does not assume LS coupling. For energies not available through NIST 
(or Kurucz) we typically used scaled theoretical energies, while hydrogenic values were adopted for high ($n, l$) states. To facilitate modeling large atoms, high $l$ states were typically grouped into a single state (denoted by $z$), while high states (e.g. \ion{C}{4}  $nl$, $n>12$ ) were treated as a single level (denoted by $w$). This grouping is done during the creation of the model atom, and is distinct from the grouping of model atom levels into super levels.

\subsection{Oscillator Strengths}

The main sources of oscillator strengths are TOPbase \citep{Topbase93}, which provides data computed by the Opacity Project \citep{Sea87_OP}, NORAD \citep{Nahar_NORAD}, and  \cite{Kur_web}. For some species, oscillator strengths have been updated with oscillator strengths from NIST. Oscillator strengths for transitions from high ($n,l$) levels are scaled hydrogenic.

Many of the calculations (e.g., those in TOPbase) assume LS coupling, and often only include levels up to a principal quantum number ($n$) of 10. We performed fine structure splitting of the oscillator strengths assuming LS coupling, and used hydrogenic vales for high $l$ and $n$ states.  

\subsection{Photoionization}

The main sources of photoionization cross-sections are the Opacity Project (\citealt{Sea87_OP}; obtained from TOPbase \citealt{Topbase93}), NORAD \citep{Nahar_NORAD}, and  \cite{Kur_web}. Hydrogenic values were also used for the photoionization cross-sections from high $l$ and $n$ states.  Recombination is treated using the same photoionization cross-sections.

\subsection{Collisional data}

For species/transitions without collisional data (or simply not yet implemented in the database) we used either the approximate formulae of \cite{1962ApJ...136..906V}, or adopted collision strengths from a nearby isoelectronic species.
 
\subsection{Species references}

\subsubsection{Carbon}

   C\,{\sc iv} oscillator strengths are from \cite{Lei72_CIV}, photoionization data is from \cite{PSS88_LI_seq} and \cite{Lei72_CIV}, and collision strengths are from \cite{CM83_lith}.
    C\,{\sc iii} oscillator strengths, photoionization data, and dielectronic line data are from P.~J. Storey (private communication). Collision rates among the lowest six terms are from \cite{BBD85_col}. C\,{\sc ii} oscillator strengths and photoionization cross-sections  are from \cite{DSK00_CII_recom} (\& P.~J.\ Storey, private communication).
   NIST \cite[][2012-10-15]{NIST_V5} values are used, when available.
    Collision rates are from \cite{Tay08_CII_col}.
    
\subsubsection{Nitrogen}
  N\,{\sc v} oscillator strengths for $n < 6$ are from \cite{LB11_Li_like},
   photoionization data is from \cite{PSS88_LI_seq},
   and collision strengths for N\,{\sc v} for $n < 6$ are from \cite{LB11_Li_like}.
    N\,{\sc iv} oscillator strengths and photoionization cross-sections were computed by \cite{TSB90_Be_seq} as part of the Opacity Project \citep{Sea87_OP}. The
      identification of the 2s\,4s\,$^3$S and 2p\,3p\,$^3$S states have been switched \citep[see][]{AAL91_NIV}. Oscillator strengths for N\,{\sc iv} forbidden (and semi-forbidden) lines are from \cite{NS79_Inter}.      N\,{\sc iv} LTDR oscillator strengths are from \cite{NS83_LTDR,NS84_CNO_LTDR}. 
     
   N\,{\sc iii} oscillator strengths and photoionization cross-sections  were computed as part of the Opacity Project \citep{Sea87_OP} and were obtained from TOPbase \citep{Topbase93}.
   Transition probabilities for intercombination lines are from \cite{NS79_NIII}. These are in reasonable agreement with experimental values found in the study of \cite{TGK99_CII_NIII}. Collision strengths for transitions amongst the first 20 levels are from \cite{SBH94_NIII_col}.
   
\subsubsection{Oxygen}
  O\,{\sc vi} oscillator strengths for $n < 6$ are from \cite{LB11_Li_like}, 
    photoionization data is from \cite{PSS88_LI_seq}, and 
   collision strengths for O\,{\sc vi} for $n < 6$ are from \cite{LB11_Li_like}.
      Oscillator strengths and photoionization cross-sections for O\,{\sc v} are from  \cite{TSB90_Be_seq} and were computed as part of the Opacity Project. Data from  \cite{NS79_Inter} were also used.   Collision rates for the six lowest terms of O\,{v} are from \cite{BBD85_col}.
   O\,{\sc iv} oscillator strengths were computed as part of the Opacity Project
   \citep{Sea87_OP}. Intercombination data ($^2$P-$^4$P) for O\,{\sc iv} is from
   the compilation of \cite{Men83_col}. Photoionization data is from \cite{Nah98_Ophot} and
   was obtained through NORAD \citep{Nahar_NORAD}, and collision rates are from \cite{ZGP94_Blike_col}.
   O\,{\sc iii} oscillator strengths and photoionization data are from \cite{LPS89_OIII_phot}.
   Collisional data for the four lowest terms in O\,{\sc iii} are from  the compilation of \cite{Men83_col}.
  
\subsection{Isoelectronic sequences} 

Lithium sequence (\ion{Ne}{8}, \ion{Na}{9}, \ion{Mg}{10}, \ion{Al}{11}):
Oscillator strengths and photoionization cross-sections are from \cite{PSS88_LI_seq}.

Beryllium sequence (\ion{Ne}{7}, \ion{Na}{8}, \ion{Mg}{9}, \ion{Al}{10}):  Oscillator strengths and photoionization cross-sections were computed by \cite{TSB90_Be_seq} as part of the Opacity Project.
            
Boron sequence (\ion{Ne}{6}, \ion{Na}{7}, \ion{Mg}{8}, \ion{Al}{9}, \ion{Si}{10}): Oscillator strengths and photoionization cross-sections were computed by J. A.~Fernley, A.~Hibbert, AE Kingston and M.~J.~Seaton (unpublished) as part of the Opacity Project. Collisional data for
the sequence is from \cite{ZGP94_Blike_col}.
  
Carbon sequence (\ion{Ne}{5}, \ion{Na}{6}, \ion{Mg}{7}, \ion{Al}{8}, \ion{Si}{9}, \ion{S}{11}): Oscillator strengths and photoionization cross-sections  are from \cite{LP89_C_seq}.
 Collisional data for the 4 lowest terms is from \cite{LB94_col}.
 
Nitrogen sequence (\ion{Ne}{4}, \ion{Na}{5}, \ion{Mg}{6}, \ion{Al}{7}, \ion{Si}{8}, \ion{S}{10}):
Oscillator strengths and photoionization cross-sections were computed by V.\ M.\ Burke and D.\ J.\ Lennon (unpublished) for the Opacity Project, and were retrieved from TOPbase. 
 
Oxygen sequence (\ion{Ne}{3}, \ion{Na}{4}, \ion{Mg}{5}, \ion{Al}{6}, \ion{Si}{7}, \ion{S}{9}): Oscillator strengths and photoionization cross-sections were computed  as part of the Opacity Project by K.\ Butler and  C.\ J.\ Zeippen (unpublished).  Intercombination rates are from \cite{GMD97_CO_Seq} and Mendoza \cite{Men83_col}. Collision rates are from  \cite{Men83_col}.

Florine sequence (\ion{Ne}{2}, \ion{Na}{3}, \ion{Mg}{4}, \ion{Al}{5}, \ion{Si}{6}, \ion{S}{8}): 
Oscillator strengths and photoionization cross-sections were computed by
     K.\ Butler and C.\ J.\ Zeippen  (unpublished)  as part the Opacity Project.
 
Neon sequence  (\ion{Mg}{3}, \ion{Al}{4}, \ion{Si}{5}, \ion{S}{7}): Oscillator strengths were computed by  K.\ Butler and C.\ J.\ Zeippen (unpublished) for Opacity Project .

Sodium sequence (\ion{Mg}{2}, \ion{Al}{3}, \ion{Si}{4}, \ion{S}{6}): Oscillator strengths and photoionization cross-sections were computed as part the Opacity Project.
Collision strengths (for $n < 7$) are from \cite{LWB09_Na_seq_col}.
 
 Ca\,{\sc vii} oscillator strengths and photoionization cross-sections were computed by
 C.\ Mendoza, W.\ Eissner, M.\ Le Dourneuf, C.\ J.\ Zeippen (unpublished) for the Opacity Project.

Ca\,{\sc vi} oscillator strengths  and photoionization cross-sections  were computed
   as part of the Opacity Project by \cite{NP93_Sil}.
 
Ca\,{\sc iv} oscillator strengths and  photoionization cross-sections were computed
    by K.\ Butler, C.\ Mendoza and C.\ J.\ Zeippen (unpublished) as part of the Opacity Project.
   
\subsubsection{Fe}

Oscillator strengths and photoionization cross-sections  for Fe\,{\sc xvi} were
computed by  K.~T.\ Taylor (unpublished) as part of the Opacity Project. 
 Oscillator strengths and photoionization cross-sections  for Fe\,{\sc xv} are
 from \cite{BKM93_Mg_seq}.
 Oscillator strengths and photoionization cross-sections for Fe\,{\sc xiv} data are from the Opacity Project
 [C.\ Mendoza, W.\ Eissner, M.\ Le Dourneuf, C.\ J.\ Zeippen (unpublished)].

Oscillator strengths and photoionization cross- sections  for Fe\,{\sc xi}, Fe\,{\sc xii}, 
and Fe\,{\sc xiii} were computed by  C.\ Mendoza (unpublished) as part of the Opacity Project.
Collision strengths for Fe\,{\sc xiii}   are from \cite{AK05_FeXIII} and
those for  Fe\,{\sc xi} are from \cite{2003A&A...399..799A}.
   
%Oscillator strengths for Fe\,{\sc ix} and Fe\,{\sc x} are from Kurucz (\citep{Kur09_ATD,Kur_web}). Photoionization cross- sections  were computed by  C. Mendoza (unpublished) as part of the Opacity Project.

%Fe\,{\sc v}, Fe\,{\sc vi} and Fe\,{\sc vii}

   Oscillator strengths for Fe\,{\sc iv} --  Fe\,{\sc x}  were computed by Bob Kurucz \citep{Kur09_ATD,Kur_web}. Photoionization cross-sections were obtained from TOPbase \citep{Topbase93}. Collision strengths for Fe\,{\sc vi} are from \cite{CP99_FeVI_col}
 while Fe\,{\sc iv} collisional data is from \cite{ZP97_FeIV_col}.

For other elements (e.g., Cl, Cr, Ni) less atomic data are available. Typically we used oscillator strengths from  Bob Kurucz \citep{Kur09_ATD,Kur_web} and NIST. We typically adopted hydrogenic photoionization cross-sections, or used cross-sections from a nearby
species in the same isoelectronic sequence. Admittedly these are crude approximations,
but they do allow more realistic models to be constructed.

\newcommand{\sr}{\rm \scriptsize}
\startlongtable
\begin{deluxetable}{lrrl}
\tablewidth{0pt}
\tablecaption{\label{tab_atom_small}Small Atomic Model}
\tablehead{
\colhead{Species}    &   \colhead{N$_{\scriptsize S}$} &   \colhead{N$_{\scriptsize F}$}   &
\colhead{Level name\tablenotemark{a}} }
\startdata
\ion{He}{1}    &     27  &     39  &   n(14)                                                 \\     
\ion{He}{2}    &     13  &     30  &   n(30)                                                     \\     
\ion{He}{3}    &      1  &      1  &                                                          \\
\ion{C}{2}     &      9  &     16  &   2s$^2$ 3d $^2$D                                        \\     
\ion{C}{3}     &    100  &    243  &   2p 4d$^1$D$^{\sr o}$                                   \\     
\ion{C}{4}     &     59  &     64  &   30                                                     \\     
\ion{C}{5}     &     43  &    107  &   2s$^2$ 30w $^1$W                                       \\     
\ion{C}{6}     &      1  &      1  &                                                          \\
\ion{N}{3}     &     34  &     71  &   2s$^2$ 5g $^2$G                                        \\     
\ion{N}{4}     &     34  &     60  &   2s 4f $^1$F$^{\sr o}$                                  \\     
\ion{N}{5}     &     45  &     67  &   n(30)                                                     \\     
\ion{N}{6}     &      1  &      1  &                                                          \\
\ion{O}{2}     &     14  &     29  &   2s$^2$ 2p$^2$($^3$P)3p $^4$S$^{\sr o}$                 \\     
\ion{O}{3}     &     85  &    267  &   2s$^2$ 2p 8g $^1$F$^{\sr o}$                           \\     
\ion{O}{4}     &    220  &    324  &   2p$^2$($^3$P)10p $^4$S$^{\sr o}$                       \\     
\ion{O}{5}     &     75  &    152  &   2s 7f $^1$F$^{\sr o}$                                   \\     
\ion{O}{6}     &     41  &     47  &   12z $^2$Z                                               \\     
\ion{O}{7}     &      1  &      1  &                                                          \\
\ion{Ne}{2}    &     42  &    242  &   2s$^2$ 2p$^4$($^1$D)4d $^2$S                           \\     
\ion{Ne}{3}    &     57  &    188  &   2s$^2$ 2p$^3$($^2$D$^{\sr o}$)5d $^1$S$^{\sr o}$       \\     
\ion{Ne}{4}    &     45  &    355  &   2s 2p$^3$($^5$S$^{\sr o}$)5s $^6$S$^{\sr o}$           \\     
\ion{Ne}{5}    &     37  &    166  &   2s$^2$ 2p 5f $^3$G                                     \\     
\ion{Ne}{6}    &     36  &    202  &   2s$^2$ 10z$^2$Z                                        \\     
\ion{Ne}{7}    &     38  &    182  &   2s 10z $^1$Z                                            \\     
\ion{Ne}{8}    &     24  &     49  &   10p $^2$P$^{\sr o}$                                    \\     
\ion{Ne}{9}    &      1  &      1  &                                                          \\
\ion{Na}{2}    &     21  &     35  &   2p$^5$($^2$P$_{\hbox{\scriptsize 1/2}}$)6s [2]$_{\hbox{\scriptsize 1/2}}^{\sr o}$                 \\    
\ion{Na}{3}    &     25  &    144  &   2p$^4$($^3$P)6f $^2$D$^{\sr o}$                        \\     
\ion{Na}{4}    &     34  &    154  &   2p$^3$($^4$S$^{\sr o}$)6f $^5$F                        \\     
\ion{Na}{5}    &     39  &    245  &   2s$^2$ 2p$^2$($^3$P)7s $^2$P                           \\     
\ion{Na}{6}    &     45  &    237  &   2s$^2$ 2p 6d $^1$P$^{\sr o}$                           \\     
\ion{Na}{7}    &     30  &    143  &   2s$^2$ 6f $^2$F$^{\sr o}$                              \\     
\ion{Na}{8}    &     72  &    214  &   2p 7d $^3$D$^{\sr o}$                                  \\     
\ion{Na}{9}    &     27  &     71  &   30w $^2$W                                              \\     
\ion{Na}{10}   &      1  &      1  &                                                          \\
\ion{Mg}{3}    &     29  &    201  &   2p$^5$ 10f $^3$F                                       \\     
\ion{Mg}{4}    &     27  &    264  &   2s 2p$^5$($^3$P$^{\sr o}$)3p $^4$S                     \\     
\ion{Mg}{5}    &     43  &    311  &   2s$^2$ 2p$^3$($^2$P$^{\sr o}$)6p $^1$D                 \\     
\ion{Mg}{6}    &     40  &    270  &   2s 2p$^3$($^1$D$^{\sr o}$)3d $^2$S$^{\sr o}$           \\     
\ion{Mg}{7}    &     47  &    260  &   2s 2p$^2$($^2$D)4p $^1$P$^{\sr o}$                     \\     
\ion{Mg}{8}    &     43  &    190  &   2s$^2$ 8g $^2$G                                        \\     
\ion{Mg}{9}    &     74  &    242  &   2s 30w $^1$W                                           \\     
\ion{Mg}{10}   &     25  &     74  &   30w $^2$W                                              \\     
\ion{Mg}{11}   &      1  &      1  &                                                          \\
\ion{Al}{3}    &     31  &     80  &   30w $^2$W                                              \\     
\ion{Al}{4}    &     41  &     97  &   2s$^2$ 2p$^5$($^2$P$^{\sr o}$)6f $^3$F                 \\     
\ion{Al}{5}    &     48  &    170  &   2s$^2$ 2p$^4$($^3$P)6d $^2$D                           \\     
\ion{Al}{6}    &     44  &    305  &   2s$^2$ 2p$^3$($^4$S$^{\sr o}$)10f $^3$F                \\     
\ion{Al}{7}    &     65  &    254  &   2s$^2$ 2p$^2$($^3$P)6f $^2$D$^{\sr o}$                 \\     
\ion{Al}{8}    &     52  &    273  &   2s$^2$ 2p 6f $^1$D                                     \\     
\ion{Al}{9}    &     34  &    135  &   2s$^2$ 6g $^2$G                                        \\     
\ion{Al}{10}   &     45  &     94  &   2s 6f $^1$F$^{\sr o}$                                  \\     
\ion{Al}{11}   &     25  &     75  &   30w $^2$W                                              \\     
\ion{Al}{12}   &      1  &      1  &                                                          \\
\ion{Si}{4}    &     37  &     48  &   10f $^2$F$^{\sr o}$                                     \\     
\ion{Si}{5}    &     33  &     71  &   2p$^5$ 5f $^3$F                                        \\     
\ion{Si}{6}    &     42  &    132  &   2s$^2$ 2p$^4$($^3$P)5d $^2$D                           \\     
\ion{Si}{7}    &     34  &    151  &   2s$^2$ 2p$^3$($^2$P$^{\sr o}$)4f $^1$G                 \\     
\ion{Si}{8}    &     37  &    194  &   2s$^2$ 2p$^2$($^3$P)5f $^2$F$^{\sr o}$                 \\     
\ion{Si}{9}    &     47  &    215  &   2s$^2$ 2p 5f $^1$D                                     \\     
\ion{Si}{10}   &     44  &    204  &   2s 2p($^1$P$^{\sr o}$)4f $^2$D                         \\     
\ion{Sk}{11}   &      1  &      1  &                                                          \\
\ion{P}{4}     &     36  &    178  &   3p 5p $^3$P                                            \\     
\ion{P}{5}     &     16  &     62  &   10g $^2$G                                              \\     
\ion{P}{6}     &      1  &      1  &                                                          \\
\ion{S}{3}     &     13  &     28  &   3s$^2$ 3p 4s $^1$P$^{\sr o}$                           \\     
\ion{S}{4}     &     51  &    142  &   3s$^2$ 8z$^2$Z                                         \\     
\ion{S}{5}     &     31  &     98  &   3p 4d $^1$P$^{\sr o}$                                  \\     
\ion{S}{6}     &     28  &     58  &   10z$^2$Z                                               \\     
\ion{S}{7}     &     36  &    207  &   2s$^2$ 2p$^5$ 10g $^3$G$^{\sr o}$                      \\     
\ion{S}{8}     &     41  &    172  &   2s$^2$ 2p$^4$($^3$P)6f $^2$D$^{\sr o}$                 \\     
\ion{S}{9}     &     44  &    233  &   2s$^2$ 2p$^3$($^4$S$^{\sr o}$)6f $^3$F                 \\     
\ion{S}{10}    &     34  &     72  &   2s$^2$ 2p$^2$($^1$S)3d $^2$D                           \\     
\ion{S}{11}    &      1  &      1  &                                                          \\
\ion{Cl}{4}    &     40  &    129  &   3p$^3$ 3d(a) $^3$D$^{\sr o}$                           \\     
\ion{Cl}{5}    &     26  &     80  &   6s $^2$S                                               \\     
\ion{Cl}{6}    &     18  &     44  &   3s 4f $^1$F$^{\sr o}$                                  \\     
\ion{Cl}{7}    &     37  &     48  &   10z $^2$Z                                               \\     
\ion{Cl}{8}    &      1  &      1  &                                                          \\
\ion{Ar}{3}    &     10  &     36  &   3s$^2$ 3p$^3$($^2$P$^{\sr o}$)3d $^3$P$^{\sr o}$       \\     
\ion{Ar}{4}    &     31  &    105  &   3s$^2$ 3p$^2$($^3$P)4f $^4$F$^{\sr o}$                 \\     
\ion{Ar}{5}    &     38  &     99  &   3s$^2$ 3p 5p $^1$P                                     \\     
\ion{Ar}{6}    &     21  &     81  &   3s$^2$ 5f $^2$F$^{\sr o}$                              \\     
\ion{Ar}{7}    &     30  &     72  &   3p 4f $^3$D                                            \\     
\ion{Ar}{8}    &     28  &     52  &   10z $^2$Z                                               \\     
\ion{Ar}{9}    &      1  &      1  &                                                          \\
\ion{Ca}{4}    &     31  &    137  &   3s$^2$ 3p$^4$($^3$P)5g $^2$F                           \\     
\ion{Ca}{5}    &     43  &    107  &   3s$^2$ 3p$^3$($^2$P$^{\sr o}$)4d $^1$F$^{\sr o}$       \\     
\ion{Ca}{6}    &     40  &    127  &   3s$^2$ 3p$^2$($^3$P)5s $^4$P                           \\     
\ion{Ca}{7}    &     48  &    288  &   3s 3p$^2$($^4$P)5p $^3$P$^{\sr o}$                     \\     
\ion{Ca}{8}    &     45  &    296  &   3s 3d 4p $^2$Fb$^{\sr o}$                              \\     
\ion{Ca}{9}    &     39  &    162  &   3s.7p $^3$P$^{\sr o}$                                  \\     
\ion{Ca}{10}   &     27  &     59  &   10f $^2$F$^{\sr o}$                                    \\     
\ion{Ca}{11}   &     32  &    150  &   2s$^2$ 2p$^5$ 6d $^1$F$^{\sr o}$                       \\     
\ion{Ca}{12}   &      1  &      1  &                                                          \\
\ion{Cr}{4}    &     29  &    234  &   3d$^2$($^3$P)5p $^4$P$^{\sr o}$                        \\     
\ion{Cr}{5}    &     30  &    223  &   3d 6p $^1$F$^{\sr o}$                                  \\     
\ion{Cr}{6}    &     30  &    215  &   3p$^5$ 3d($^1$P)4s $^2$P$^{\sr o}$                     \\     
\ion{Cr}{7}    &      1  &      1  &                                                          \\
\ion{Mn}{4}    &     39  &    464  &   3d$^3$($^4$P)5p $^5$S$^{\sr o}$                        \\     
\ion{Mn}{5}    &     16  &     80  &   3d$^2$($^1$S)4p $^2$P$^{\sr o}$                        \\     
\ion{Mn}{6}    &     23  &    181  &   3d 4f $^3$Pb$^{\sr o}$                                 \\     
\ion{Mn}{7}    &     20  &    203  &   10p $^2$P$^{\sr o}$                                    \\     
\ion{Mn}{8}    &      1  &      1  &                                                          \\
\ion{Fe}{4}    &     53  &    314  &   3d$^4$($^1$D1)4p $^2$D$^{\sr o}$                       \\     
\ion{Fe}{5}    &     61  &    300  &   3d$^3$($^4$F)5s $^5$F                                  \\     
\ion{Fe}{6}    &     44  &    433  &   3p$^5$($^2$P)3d$^4$($^1$S) $^2$Pc$^{\sr o}$            \\     
\ion{Fe}{7}    &     29  &    153  &   3p$^5$($^2$P)3d$^3$(b$^2$D) $^1$P$^{\sr o}$            \\     
\ion{Fe}{8}    &     53  &    324  &   3p$^5$3d(b$^3$D)4s$^4$D$^{\sr o}$                      \\     
\ion{Fe}{9}    &     52  &    490  &   ($^3$P)3d($^4$P)5p $^3$S$^{\sr o}$                    \\     
\ion{Fe}{10}   &     43  &    210  &   3s$^2$ 3p$^2$ 3d$^3$(b) $^4$P                          \\     
\ion{Fe}{11}   &     72  &   1154  &   3s 3p$^4$($^4$P)7d $^5$D                               \\     
\ion{Fe}{12}   &     69  &    915  &   3s 3p$^3$($^5$S$^{\sr o}$)9f $^6$F                     \\     
\ion{Fe}{13}   &     72  &    819  &   3s 3p$^2$($^4$P)9g $^5$G                               \\     
\ion{Fe}{14}   &     48  &    669  &   3s 3p($^1$P$^{\sr o}$)8f $^2$F                         \\     
\ion{Fe}{15}   &      1  &      1  &                                                          \\
\ion{Ni}{4}    &     36  &    200  &   3d$^6$($^3$D)4p $^2$D$^{\sr o}$                        \\     
\ion{Ni}{5}    &     46  &    183  &   3d$^5$($^2$D3)4p $^3$F$^{\sr o}$                       \\     
\ion{Ni}{6}    &     37  &    314  &   3d$^4$($^5$D)4d $^4$F                                  \\     
\ion{Ni}{7}    &     37  &    308  &   3d$^3$($^2$D)4d $^3$P                                  \\     
\ion{Ni}{8}    &     34  &    325  &   3d$^2$($^1$G)4d $^2$D                                  \\     
\ion{Ni}{9}    &     34  &    363  &   3p$^5$($^2$P)3d$^2$($^1$D)4p $^1$F                     \\     
\ion{Ni}{10}   &      1  &      1  &                                                          \\
\enddata
\tablenotetext{a}{The level name given here is the highest level included.  Generally LS naming is used, but in some case identifiers have been added to distinguish levels
with identical names. W states are used for states
summed over all $l$ -levels, but different multiplicities
are kept separate. Z states are high $l$ stats
(e.g, g and higher) that have been combined into a single
level in the model ion. In the model atoms high
level terms are generally treated as single level,
but lower terms are split.}
\end{deluxetable}

\begin{deluxetable}{lrrrl}
\tablewidth{0pt}
\tablecaption{\label{tab:small_atom}Comparison between small and intermediate atomic models}
\label{tab_atom_im}
\tablehead{
\colhead{Species}       
& \colhead{N$_{\scriptsize S}$/N$_{\scriptsize F}$}   & \colhead{Level name} 
& \colhead{N$_{\scriptsize S}$/N$_{\scriptsize F}$}   & \colhead{Level name} \\
& \multicolumn{2}{c}{``Small" model}
& \multicolumn{2}{c}{``Intermediate" model}
}
\startdata
\ion{C}{5}     &     43/107  &   2s$^2$ 30w $^1$W  &     43/107  &   1s 30w $^1$W                                 \\
\ion{C}{6}     &      1/1    &                     &     30/75   &   30w $^2$W                                    \\
\ion{C}{7}     &             &                     &     1/1    &                                                 \\
\ion{O}{5}     &     75/152  &   2s 7f$^1$F$^{\sr o}$  &  82/234  &   2s 30w $^1$W                                  \\
\ion{O}{6}     &     41/47  &   12z$^2$Z           &     59/ 65  &   30                                           \\
\ion{Ne}{6}    &     36/202  &   2s$^2$ 10z$^2$Z   &     59/239  &   2s 2p($^3$P$^{\sr o}$)6d $^4$P$^{\sr o}$      \\
\ion{Na}{3}    &     25/144  &   2p$^4$($^3$P)6f $^2$D$^{\sr o}$   &   25/145  &   2p$^4$($^3$P)6f $^2$D$^{\sr o}$      \\
\ion{Mg}{4}    &     27/ 264  &   2s 2p$^5$($^3$P$^{\sr o}$)3p $^4$S  &     27/264  &   2s$^2$ 2p$^4$($^1$D)7d $^2$S                           \\
\ion{Al}{7}    &     65/254  &   2s$^2$ 2p$^2$($^3$P)6f $^2$D$^{\sr o}$  &    68/289  &   2s$^2$ 2p$^2$($^3$P)7p $^4$S$^{\sr o}$   \\
\ion{Fe}{9}    &     52/490  &   ($^3$P)3d($^4$P)5p $^3$S$^{\sr o}$     &     59/712  &   3s$^2$ 3p$^5$($^2$P)9k $^3$Ib       \\
\ion{Fe}{10}   &     43/210  &   3s$^2$ 3p$^2$ 3d$^3$(b) $^4$P    &    115/1368  &   3s 3p$^5$($^3$P$^{\sr o}$)5f $^2$F        \\
\ion{Fe}{11}   &     72/1154  &  3s 3p$^4$($^4$P)7d $^5$D     &    118/1236  &   3s 3p$^4$($^4$P)7f $^5$F$^{\sr o}$            \\
\ion{Fe}{12}   &     69/915  &   3s 3p$^3$($^5$S$^{\sr o}$)9f $^6$F  &    144/912  &   3s 3p 3d$^3$(a) $^2$P$^{\sr o}$         \\
\ion{Fe}{13}   &     72/819  &   3s 3p$^2$($^4$P)9g $^5$G        &    115/1239  &   3s 3p$^2$($^4$P)9g $^5$G                   \\
\ion{Fe}{14}   &     48/669  &   3s 3p($^1$P$^{\sr o}$)8f $^2$F  &     64/739  &   3s$^2$ 7h $^2$H$^{\sr o}$                   \\
\ion{Fe}{15}   &      1/ 1  &                                    &     64/448  &   3d 6g $^3$F                                 \\
\ion{Fe}{16}   &            &                                    &     20/77  &   2p$^6$ 30w $^2$W                             \\
\ion{Fe}{17}   &            &                                    &      1/1   &                                               \\
\ion{Ni}{5}    &     46/183  &   3d$^5$($^2$D3)4p $^3$F$^{\sr o}$      &    152/1000  &   3d$^4$($^1$I)4s$^2$ $^1$I                \\
\ion{Ni}{6}    &     37/314  &   3d$^4$($^5$D)4d $^4$F                 &     62/1000  &   3d$^4$($^3$D)4f $^2$Gb$^{\sr o}$        \\
\ion{Ni}{7}    &     37/308  &   3d$^3$($^2$D)4d $^3$P                 &     61/1000 &   3d$^3$($^4$F)6p $^3$F$^{\sr o}$         \\
\ion{Ni}{8}    &     34/325  &   3d$^2$($^1$G)4d $^2$D                 &     48/ 1000  &   3p$^5$($^2$P)3d$^3$($^2$H)4s $^4$G$^{\sr o}$           \\
\ion{Ni}{10}   &      1/1    &                                         &     56/324  &   3p$^5$ 3d(b$^3$D)4s $^4$D$^{\sr o}$            \\
\ion{Ni}{11}   &      1/1    &                                          &     89/499  &   3s$^2$ 3p$^5$($^2$P)6h $^3$Gb                 \\
\ion{Ni}{12}   &      1/1    &                                         &     46/233  &   3s$^2$ 3p$^4$($^3$P)4f $^2$D$^{\sr o}$         \\
\ion{Ni}{13}   &      1/1    &                                          &     73/1236  &   3s 3p$^4$($^4$P)7f $^5$F$^{\sr o}$    \\
\ion{Ni}{14}   &      1/1    &                                           &     80/912  &   3s 3p 3d$^3$(a) $^2$P$^{\sr o}$       \\
\ion{Ni}{15}   &      1/1    &                                          &    122/1239  &   3s 3p$^2$($^4$P)9g $^5$G              \\
\ion{Ni}{16}   &      1/1    &                                          &     48/669  &   3s 3p($^1$P$^{\sr o}$)8f $^2$F         \\
\ion{Ni}{17}   &             &                                          &    1/1      &              \\
\enddata
\end{deluxetable}

\begin{deluxetable}{lrrl}
\tablewidth{0pt}
\tablecaption{\label{tab:large_atom}Additional ions in largest model}
\label{tab_atom_large}
\tablehead{
\colhead{Species}       
& \colhead{N$_{\scriptsize S}$} & \colhead{N$_{\scriptsize F}$}   & \colhead{Level name} 
}
\startdata
\ion{Ar}{9}    &     26  &    247  &   2s$^2$ 2p$^5$ 10g $^3$G$^{\sr o}$                      \\  
\ion{Ar}{10}   &     32  &    315  &   2s$^2$ 2p$^4$($^1$S)8s $^2$S                           \\  
\ion{Ar}{11}   &      1  &      1  &                                                          \\
\ion{K}{3}     &     20  &     40  &   3p$^4$($^1$S)3d $^2$D                                  \\  
\ion{K}{4}     &     40  &    612  &   3p$^3$($^2$D$^{\sr o}$)10h $^1$G                       \\  
\ion{K}{5}     &     36  &    305  &   3s 3p$^3$($^3$P)4d $^2$D$^{\sr o}$                     \\  
\ion{K}{6}     &     54  &    513  &   3s 3p$^2$($^4$P)7p $^5$S$^{\sr o}$                     \\  
\ion{K}{7}     &     48  &    205  &   3s 3p($^3$P$^{\sr o}$)6d $^4$P$^{\sr o}$               \\  
\ion{K}{8}     &     35  &     85  &   3d 4s $^3$D                                            \\  
\ion{K}{9}     &     22  &     72  &   30w $^2$W                                               \\  
\ion{K}{10}    &     27  &    247  &   2s$^2$ 2p$^5$ 10g $^1$G$^{\sr o}$                      \\  
\ion{K}{11}    &     32  &    315  &   2s$^2$ 2p$^4$($^1$S)8s $^2$S                           \\  
\ion{K}{12}    &      1  &      1  &                                                          \\
\ion{Cr}{7}    &     52  &    712  &   3s$^2$ 3p$^5$($^2$P)9k $^3$Ib                          \\  
\ion{Cr}{8}    &     43  &   1368  &   3s 3p$^5$($^3$P$^{\sr o}$)5f $^2$F                     \\  
\ion{Cr}{9}    &    109  &   1236  &   3s 3p$^4$($^4$P)7f $^5$F$^{\sr o}$                     \\  
\ion{Cr}{10}   &     71  &    912  &   3s 3p 3d$^3$(a) $^2$P$^{\sr o}$                        \\  
\ion{Cr}{11}   &    113  &   1239  &   3s 3p$^2$($^4$P)9g $^5$G                               \\   
\ion{Cr}{12}   &     61  &    739  &   3s$^2$ 7h $^2$H$^{\sr o}$                              \\   
\ion{Cr}{13}   &     62  &    448  &   3d 6g $^3$F                                            \\   
\ion{Cr}{14}   &     21  &     77  &   2p$^6$ 30w $^2$W                                       \\   
\ion{Cr}{15}   &      1  &      1  &                                                          \\
\ion{Ni}{9}    &     48  &   1217  &   4s 4d  $^1$D                                     \\   
\enddata
\end{deluxetable}

\bibliography{WCPaper}{}
\bibliographystyle{aasjournal}

\end{document}